 \documentclass[final,3p,times]{elsarticle}

\usepackage{graphicx}
\usepackage{subfig}
\usepackage{caption}
\usepackage{psfrag}  

\usepackage{amssymb}
\usepackage{graphicx}
\usepackage{amsmath}%
\usepackage[normalem]{ulem} 
\usepackage{wasysym}%
\usepackage{subfig}
\usepackage{psfrag}
\usepackage[usenames, dvipsnames]{color}
\usepackage{float}
\usepackage{subfloat}
\usepackage{natbib}
\usepackage{pifont}
\usepackage{listings}
\usepackage{cases}
\usepackage{enumerate}
\usepackage{tabulary}

\newcommand\Rey{\mbox{\textit{Re}}}  
\newcommand\Pran{\mbox{\textit{Pr}}} 

\journal{Journal of Computational Physics}

\begin{document}

\begin{frontmatter}

\title{An efficient semi-implicit solver for direct numerical simulation \\ of compressible flows at all speeds}

\author{Davide Modesti~$^\ast$ and Sergio Pirozzoli}
\ead{davide.modesti@uniroma1.it}
\cortext[cor1]{Corresponding author. Tel.  +39-06-44585202, Fax +39-06-44585250}
\address{Sapienza Universit\`a di Roma, Dipartimento di Ingegneria Meccanica e Aerospaziale, via Eudossiana 18, 00184 Roma, Italy}

\begin{abstract}
We develop a semi-implicit algorithm for time-accurate simulation of the compressible Navier-Stokes equations,
with special reference to wall-bounded flows. The method is based on linearization of the partial convective fluxes 
associated with acoustic waves, in such a way to suppress, or at least mitigate the acoustic time step limitation.
Together with replacement of the total energy equation with the entropy transport equation, this approach
avoids the inversion of block-banded matrices involved in classical methods, which is replaced by less demanding inversion
of standard banded matrices. The method is extended to deal with implicit integration of viscous terms
and to multiple space dimensions through approximate factorization, and used as a building
block of third-order Runge-Kutta time stepping scheme. 
Numerical experiments are carried out for isotropic turbulence, plane channel flow, and flow in a square duct.
All available data support higher computational efficiency than existing methods, and saving of
resources ranging from $85\%$ under low-subsonic flow conditions, to about $50\%$ in supersonic flow.
\end{abstract}

\end{frontmatter}

\section{Introduction}

Compressible wall-bounded flows play an important role in many
aerospace applications of industrial and academic interest.
The direct numerical solution (DNS) of the compressible Navier-Stokes 
equations for wall-bounded turbulent flows has recently become affordable
owing to the large increase in available computer power, and canonical incompressible flows
have been simulated up to high Reynolds number~\citep{lee_15}.
However, it is know that the numerical solution of the compressible 
Navier-Stokes equations is significantly more time consuming
than their incompressible counterpart, partly
owing to the inherently higher number of floating point operations (flops) 
per grid point, but mainly because of the much smaller time step imposed by
the acoustic stability restriction.
In free-shear flows, conventional explicit algorithms can still be used efficiently 
as long as the typical Mach number is of the order of unity.
However, wall-bounded flows inevitably include regions 
with near stagnant flow and tiny grid spacing adjacent to solid surfaces,
which makes the acoustic time step limitation
in the wall-normal direction dominant, even at high bulk Mach numbers.
Besides being dictated by stability considerations, 
time step limitations in turbulent flows also have a physical 
interpretation, as in order to capture the relevant physics of transport phenomena
with given speed (say $U$) on a mesh with given size (say $\Delta$),
time steps no larger than $\Delta/U$ should be used.
Hence, CFL numbers (defined as the ratio of the time advancement step
to the maximum allowed time step for explicit time integration) should
always be of the order of unity for genuine DNS. 
In compressible flows, information simultaneously propagate at the hydrodynamic 
and at the acoustic speed.
However, acoustic waves typically make a negligible contribution to the
overall energetics of turbulent flows~\citep{lele_94}. Hence, with the obvious exception
of cases where acoustic instabilities play an important role, such as in certain 
combustion applications~\citep{poinsot_87} or in direct simulation of aerodynamic noise~\citep{colonius_04}, 
using a time step which allows to resolve the hydrodynamic 
(vortical) mode while giving up accurate representation of acoustic phenomena
may be a legitimate choice, which actually subtends much of the research carried out 
for low-speed solvers.

It is the goal of this paper to develop a numerical algorithm for 
direct numerical simulation of compressible 
flow which is capable of seamless efficient operation throughout the 
Mach number range, down to nearly incompressible conditions.
The algorithm is at the same time meant to remove or at least alleviate the acoustic time
step limitation in the presence of solid boundaries.
To gain a clearer perception for the problem, we refer to a canonical compressible 
boundary layer flow over a flat surface, or flow in a planar channel.
Let $\Delta x$, $\Delta z$ be the mesh
spacings in the streamwise and spanwise directions, respectively, and let
$\Delta y$ be the minimum mesh spacing in the wall-normal direction, assuming unit CFL number,
the time step limitations associated with the discretization of the convective terms in the coordinate directions are
\begin{equation}
\begin{array}{ccl}
\Delta t_x^+&=&\frac{\Delta x^+}{\max(u_0^++c_0^+,c_w^+)}= \Delta x^+ M_0 \sqrt{{C_f}/{2}} \min \left(1, \frac 1{1+M_0} \sqrt{{T_w}/{T_0}} \right) , \\
\Delta t_y^+&=&\frac{\Delta y^+}{c_w^+}= \Delta y^+ M_0 \sqrt{{C_f}/{2}} \\
\Delta t_z^+&=&\frac{\Delta z^+}{\max(c_0^+,c_w^+)}= {\Delta z^+} M_0 \sqrt{{C_f}/{2}} \min \left( 1, \sqrt{{T_w}/{T_0}} \right), 
\end{array}
\label{eq:dtinv}
\end{equation}
where the `+' superscript is used to denote quantities made nondimensional with respect to local wall units, namely 
the friction velocity $u_{\tau}=(\tau_w/\rho_w)^{1/2}$, and the viscous length scale $\delta_v=\nu_w/u_{\tau}$,
the subscript $0$ is used to denote flow properties at the centerline (for channels) and at
the free-stream (for boundary layers), and $w$ to denote wall properties, with
$C_f=2 \tau_w/(\rho_0 u_0^2)$. It should be noted that if acoustic waves are suppressed, 
as is the case of strictly incompressible flow, the time step is controlled by the 
streamwise direction, and
\begin{equation}
\Delta t_I^+={\Delta x^+} \sqrt{C_f/2}. \label{eq:dtinc}
\end{equation}
The viscous time step limitation is mainly effective in the wall-normal direction, 
and in wall units one has 
\begin{equation}
\Delta t_{yv}^+= {\Delta y^+}^2.
\label{eq:dtvis}
\end{equation}
\begin{figure}
 \centering
 \psfrag{a}[t][][1.0]{}
 \psfrag{b}[t][][1.0]{}
 \psfrag{x}[t][][1.0]{$M_0$}
 \psfrag{y}[b][][1.0]{$\Delta t^+/\sqrt{C_f/2}$}
 (a)
 \includegraphics[width=3.8cm,clip,angle=270]{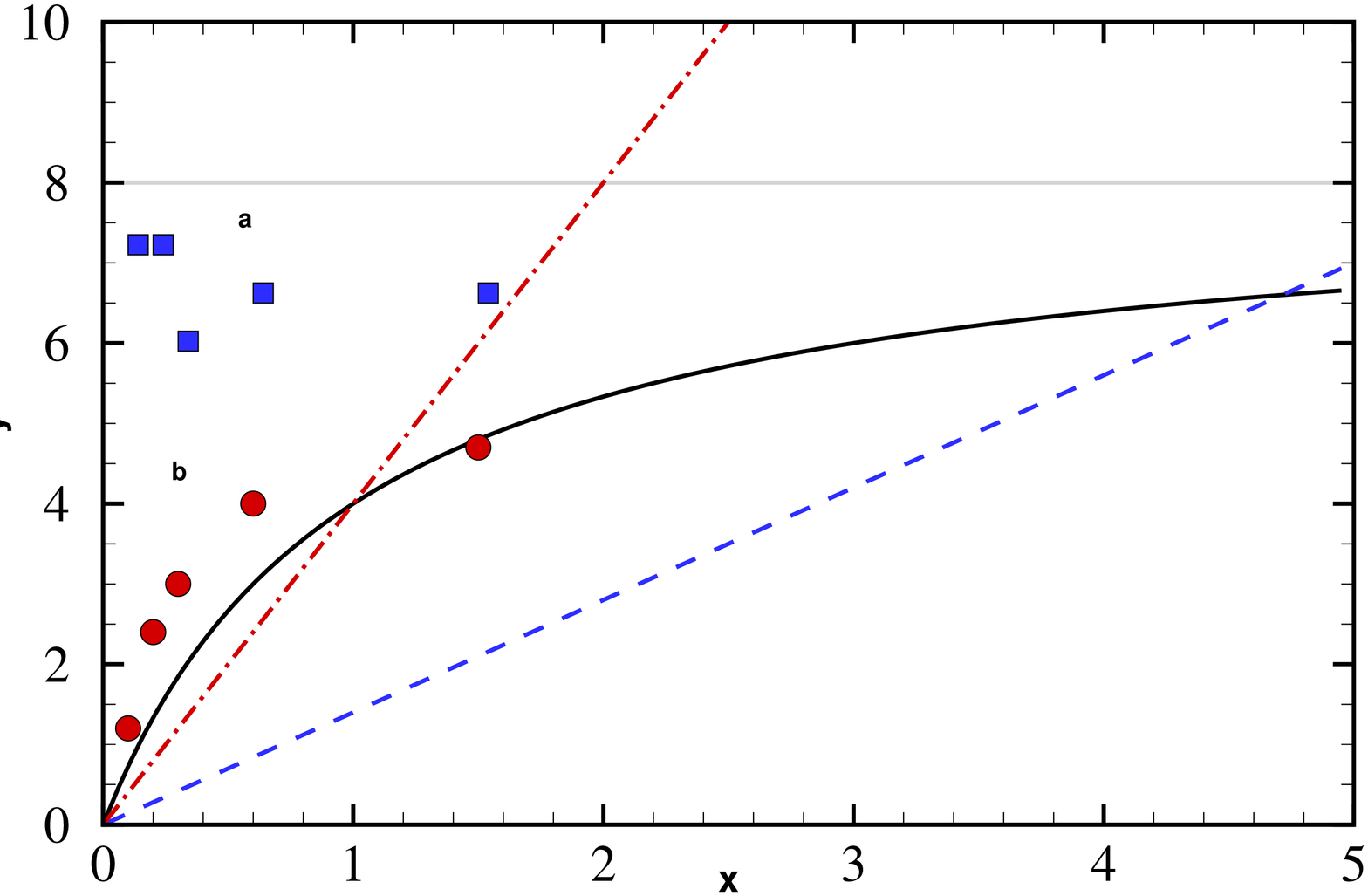} \hskip1.em
 (b)
 \psfrag{y}[b][][1.0]{$\Delta t_i^+/\Delta t_y^+$}
 \includegraphics[width=3.8cm,clip,angle=270]{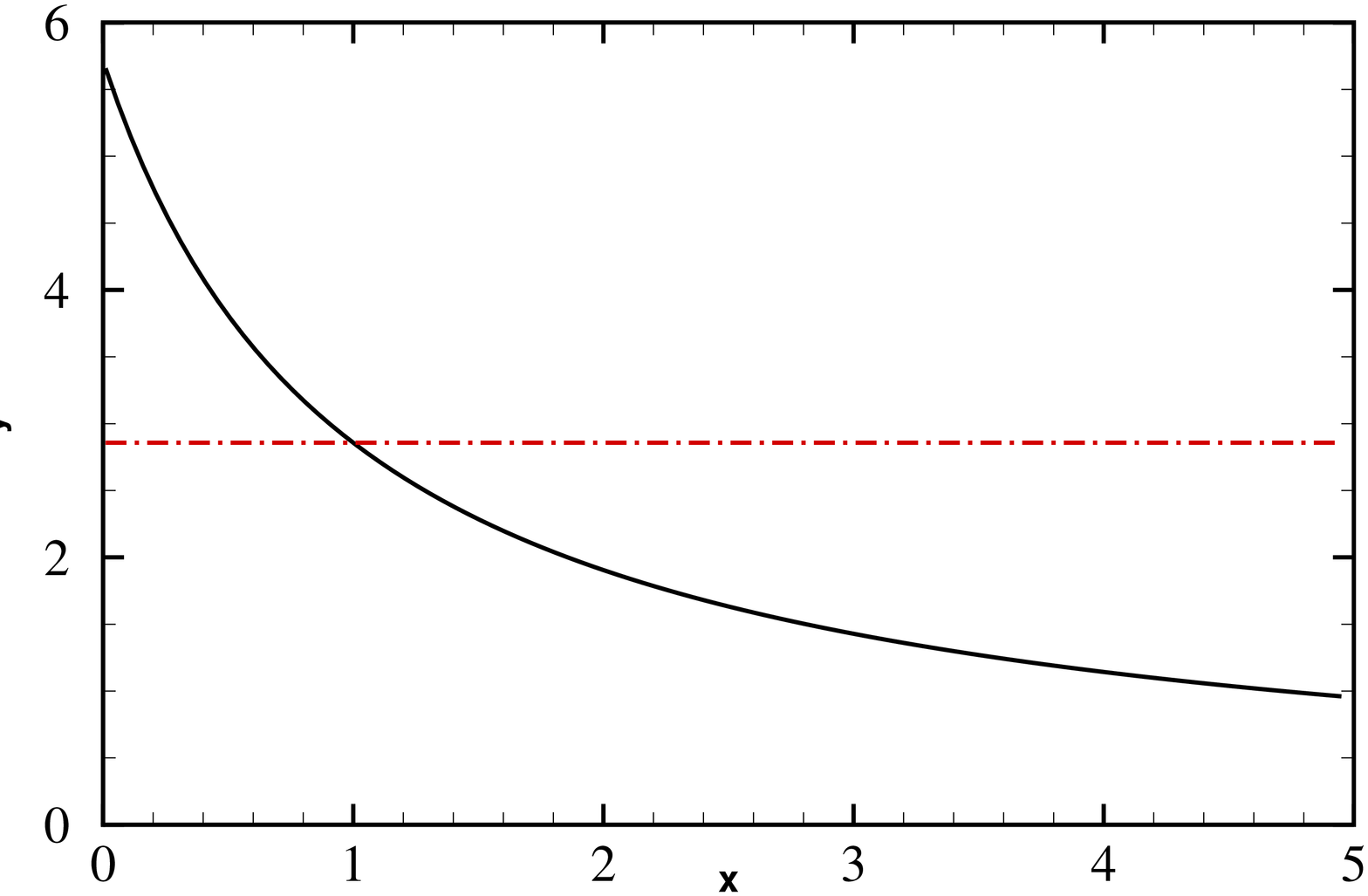}
 \caption{Inviscid time step limitation in the coordinate directions as from Eqn.~\eqref{eq:dtinv} as a function of the reference Mach number $M_0$. In panel (a) we show $\Delta t_x$ (solid), $\Delta t_y$ (dashed), $\Delta t_z$ (dot-dashed). In panel (b) we show the ratios $\Delta t_x/\Delta t_y$ (solid), $\Delta t_z/\Delta t_y$ (dot-dashed). For reference, in panel (a) we report with a grey line the `incompressible' time limitation given in Eqn.~\eqref{eq:dtinc}. The symbols denote the time step limits for the ATI algorithm as dictated by accuracy (circles) and stability (squares), as discussed in Section~\ref{sec:channel}.}
 \label{fig:dtconv}
\end{figure}
For the sake of graphical representation of the above formulas, we assume:
i) the distance of the first point from the wall is $\Delta y_w^+ \approx 0.7$,
which is the maximum value for which accurate turbulence statistics are obtained~\citep{modesti_16};
ii) the minimum mesh spacing in the wall-normal direction is $\Delta y = 2 \Delta y_w$,
which can be achieved by staggering the mesh in the vertical direction, thus alleviating the
stability restrictions~\citep{modesti_16};
iii) the wall-parallel mesh spacings are $\Delta x^+ = 8$, $\Delta z^+ = 4$,
which is typical for DNS;
 iv) the wall is isothermal, with $T_w=T_0$.
Figure~\ref{fig:dtconv} shows the inviscid time step restrictions according to Eqn.~\eqref{eq:dtinv}
as a function of the reference Mach number $M_0$, scaled by $\sqrt{C_f/2}$ (panel a),
and as a fraction of the wall-normal allowed time step (panel b). 
Inefficiency of explicit compressible solvers is apparent 
in the low-Mach-number regime, where vanishingly small time steps are required.
Time steps comparable to those achievable in incompressible flow
are only possible starting at $M_0 \approx 3$. With the exception of hypersonic flow, the most restrictive 
time limitation is that associated with the vertical direction, and an increase by at least a factor of two can 
be gained by removing it (see panel b). It is also interesting to note that the acoustic
time limitation in the spanwise size is more restrictive than the streamwise limitation up to $M_0 \approx 1$,
whereas at supersonic Mach numbers the convective limitation in $x$ is controlling.
Removing the wall-normal acoustic time limitation in supersonic flow 
is sufficient to achieve a similar time step as in incompressible flow, 
whereas in subsonic flow it is also necessary to remove the acoustic time restriction in
the wall-parallel directions.
We further note that the normalized viscous time limitation $\Delta t_{yv}^+/\sqrt{C_f/2}$, with $\Delta t_{yv}^+$ given
in Eqn.~\eqref{eq:dtvis} is always much weaker than the convective ones, 
provided $\Delta y^+ \sim 1$, and considering that the range of friction coefficients
typically accessed by DNS is $2 \times 10^{-3} \le C_f \le 6 \times 10^{-3}$.
While the above estimates are reported for typical DNS mesh spacings, the case of 
wall-resolved RANS, LES and DES is even more severe, as the aspect ratio of near-wall cells
is substantially higher, hence making suppression of the wall-normal time step restriction 
mandatory for any practical calculation.

All the above-mentioned difficulties are well know to the CFD community, and 
a variety of techniques have been developed to cope with the numerical 
stiffness of the compressible Navier-Stokes equations.
The chief choice in this respect has traditionally been the use
of (semi-)implicit time integration schemes.
A landmark contribution in this sense was given by \citet{beam_76,beam_78},
who proposed a time-implicit algorithm for the solution of the Navier-Stokes equations
in conservative form based on linearization of the convective and viscous flux vectors,
coupled with approximate factorization~\citep{douglas_55} to handle multiple space dimensions.
However, the method is computationally expensive as it requires the inversion of
$5 \times 5$ block-banded systems of equations, which is more expensive than,
e.g. standard banded systems.
In this respect we note that, whereas the classical Thomas algorithm for tridiagonal
matrices requires a number of floating point operations (flops) of $O(6N)$ (where $N$ is the number of grid
points in a given coordinate direction), its block-tridiagonal version requires $O(3N(M^3+M^2))$ flops,
where $M$ ($=5$ in the Beam-Warming algorithm) is the size of each block~\citep{isaacson_94}.
The computational cost is about twice as much in the case of periodic boundary conditions~\citep{batista_06}.
\citet{pulliam_81} developed a variant of the Beam-Warming algorithm which involves the inversion
of standard tridiagonal systems rather than block matrices, with large saving of computer 
time, but with loss of accuracy and stability in the
case of unsteady simulations~\citep{hirsch_07}.
Algorithms of the Beam-Warming family are at the heart of highly successful 
aerospace CFD software~\citep{pulliam_86,buning_98}.
Algorithms which avoid inversion of banded systems of equations
have also been designed~\citep{martin_06}, which may be useful
for efficient parallel implementation.
However, those algorithms require point-wise iterative procedures whereby the
right-hand-side of the equations must be evaluated several times per time step,
with unclear outcome in terms of overall efficiency.

Alternative approaches to circumvent the stiffness of compressible Navier-Stokes equations
rely on the use of pre-conditioning techniques, based on the attempt to
change the eigenvalues of the system of equations in order
to remove the large disparity of wave speeds. This is accomplished by 
pre-multiplying the time derivatives by a matrix that slows the speed of the acoustic
waves down toward the fluid speed~\citep{turkel_93,turkel_99}.
Preconditioning is the choice of election for steady-state application,
however its extension to unsteady flow problem is not straightforward, 
requiring the use of dual time stepping techniques, 
namely inner iterations in terms of a pseudo-time~\citep{venkateswaran_95,pandya_03,depalma_06}.
However, the number of iterations per physical
time step can be very large, with subsequent loss of computational efficiency.

Specialized algorithms for the Navier-Stokes equations
have been also developed for the low-Mach number regime, 
which allow to account for temperature-dependent density variations,
as is typically the case in combustion.
All these variable-density algorithms are based on the idea the only the terms which
bring an acoustic contribution should be advanced implicitly in time, in
such a way that the acoustic time limitation is removed.
Numerical schemes of this kind were pioneered by 
\citet{casulli_84}, who proposed to treat implicitly only
the pressure term in the momentum equation and the dilatation term
in the internal energy equation, which results in having to solve an elliptic 
equation for pressure, with large incurred overhead.
\citet{pierce_01,wall_02} extended the classical pressure-correction method~\citep{kim_85} 
to variable-density flows by
solving a Helmholtz equation for the pressure correction, and the use of sub-iterations.
LES results were carried out in which a time step forty times larger than the explicit
case was achieved, with modest computational cost overhead.
\citet{moureau_07} developed
an implicit scheme for the removal of the acoustic limitation
which also relies on the solution of a Helmholtz equation, however without 
reverting to sub-iterations, with an overhead CPU time of about $25\%$
with respect to standard incompressible solvers.
Hence it appears that, in one way or another, algorithms tailored for the near-incompressible
regime involve either iterative procedures and/or the inversion of elliptic systems of equations.
The latter can only be carried out efficiently in the case that periodic directions
are present, which allows for the use of direct solvers~\citep{spalart_91}.
 
In this paper we develop a novel semi-implicit algorithm for the compressible Navier-Stokes equations 
based on a modification of the basic Beam-Warming linearization, thus avoiding any iterative procedure.
The algorithm is presented in Section~\ref{sec:numerics}, which also includes a discussion of
the treatment of viscous terms, accurate time integration, and extension to multiple space dimensions.
Numerical examples are given in Section~\ref{sec:results}, which include DNS of turbulent flows from
the low subsonic to the supersonic regime. 
Final remarks and suggestions for future work are given in Section~\ref{sec:conclusions}.

\section{Formulation of the algorithm} \label{sec:numerics}

The Navier-Stokes equations for a compressible perfect gas are considered 
in which the total energy equation is replaced with the entropy equation 
\begin{equation}
 \frac{\partial \mathbf{w}}{\partial t}=-\sum_{i=1}^3\frac{\partial\mathbf{f}_i}{\partial x_i}+\sum_{i=1}^3\frac{\partial \mathbf{f}_i^v}{\partial x_i} + \mathbf{S} = \mathbf{R} ,
\label{eq:NS}
\end{equation}
where $\mathbf{w}$ is the vector of the conserved variables,
$\mathbf{f}_i$ and $\mathbf{f}_i^v$ are the convective and
viscous fluxes in the $i$-th direction, with $x,y,z$ the
streamwise, wall normal and spanwise directions and $\mathbf{S}$ the source terms
in the entropy equation, 
\begin{equation}
\mathbf{w}=
\begin{bmatrix}
 \rho\\ \rho u_j\\\rho s
\end{bmatrix}, \quad
\mathbf{f}_i=
\begin{bmatrix}
 \rho u_i\\ \rho u_iu_j + p\delta_{ij}\\\rho u_i s
\end{bmatrix}, \quad
\mathbf{f}_i^v=
\begin{bmatrix}
 0 \\ \sigma_{ij}\\ -{q_i}/{T}
\end{bmatrix}, \quad
\mathbf{S}=
\begin{bmatrix}
 0 \\ 0 \\ 0 \\ 0 \\ \frac{\sigma_{\ell m}}{T}\frac{\partial u_{\ell}}{\partial x_m}
  -\frac{q_{\ell}}{T^2}\frac{\partial T}{\partial x_{\ell}}
\end{bmatrix},
\end{equation}
where $\rho$ is the density, p is the pressure, T is the temperature and $u_i,\,i=1,2,3$ the velocity components in the 
$i$-th direction (also denoted as $u,v,w$ in the following),
$s=c_v\ln{(p\rho^{-\gamma})}$ is the entropy per unit mass,
$q_i$ and $\sigma_{ij}$ are the components of the viscous stress tensor and heat flux,
\begin{equation}
 \sigma_{ij}=\mu\left(\frac{\partial u_i}{\partial x_j} + \frac{\partial u_j}{\partial x_i} -\frac{2}{3} \frac{\partial u_k}{\partial x_k} \delta_{ij}\right), \quad 
 q_i=-k\frac{\partial T}{\partial x_i}, 
\label{eq:heat_vec}
\end{equation}
where $\mu$ is the dynamic viscosity, $k=\mu c_p/\Pran$ the thermal conductivity and $\Pran=0.72$ the molecular Prandtl number.

As shown in the following, the use of the entropy equation is instrumental to achieving efficient 
implicit treatment of the acoustic terms, and also yield benefits in terms of 
increased robustness as compared to algorithms solving for the energy equation~\citep{sesterhenn_00,honein_04}.
On the other hand, this setting prevents correct capturing of shock waves~\citep{salas_96}, 
hence in the following we restrict ourselves to discussing the case
of smooth compressible flows. Possible extensions to shocked flows will be discussed in the Conclusions.

\subsection{Implicit treatment of acoustic waves}\label{sec:beam}

In order to remove the time acoustic time step limitation in the generic coordinate direction (say, $y$),
we proceed by splitting the convective flux vector into a purely advective part, and a part
which supports acoustic fluctuations, namely 
\begin{equation}
 \mathbf{f}_y = \mathbf{f}_y^c + \mathbf{f}_y^a, \quad  
 \mathbf{f}_y^c = \begin{bmatrix} 0 \\ \rho u v \\ \rho v^2 \\ \rho v w \\ \rho v s \end{bmatrix}, \quad
 \mathbf{f}_y^a = \begin{bmatrix} \rho v \\ 0 \\ p \\ 0 \\ 0 \end{bmatrix} .
\end{equation}
In a linearized setting, this splitting yields full decoupling of
the acoustic, vortical and entropy modes~\citep{kovasznay_53}.
The main advantage for numerical purposes is that the acoustic partial flux Jacobian
has a simple structure,
\begin{equation}
\mathbf{A}_y^a = \frac{\partial{\mathbf{f}_y^a}}{\partial \mathbf{w}} =
\begin{bmatrix} 0& 0 &1& 0& 0\\
                             0& 0& 0& 0& 0&\\
 \frac{p}{\rho}\left(\gamma-\frac{s}{C_v}\right)&0&0&0&\frac{p}{\rho C_v}\\
                             0& 0& 0& 0& 0&\\
                             0& 0& 0& 0& 0&\\
\end{bmatrix} .
\label{eq:acoustic_jac}
\end{equation}
Splitting of the flux vectors into pressure and velocity contributions was previously
considered by \citet{steger_78,barth_85}, based on the attempt to reduce the block size
in the implicit operator as compared to the Beam-Warming algorithm. 
In essence, these decompositions amounted~\citep{pulliam_86} to isolating the 
pressure gradient in the momentum equation and the pressure flux in the total energy equation.
However, besides being consistent with wave decomposition in a linear setting, we find 
the splitting~\eqref{eq:acoustic_jac} to be vastly more robust in practice.

We proceed to discretize Eqn.~\eqref{eq:NS} 
between two consecutive time levels $n$ and $n+1$, by evaluating explicitly the advective partial flux,
and evaluating the acoustic partial flux
implicitly, upon linearization about time level $n$, namely
\begin{equation}
{\mathbf{f}_y^a}^{n+1} = {\mathbf{f}_y^a}^{n} + {\mathbf{A}_y^a}^n \left( \mathbf{w}^{n+1} - \mathbf{w}^{n} \right) + O(\Delta t^2),
\end{equation}
thus obtaining
\begin{equation}
\left( \mathbf{I} + \Delta t \frac{\partial}{\partial y} {\mathbf{A}_y^a}^n \right) \Delta \mathbf{w}^{n}
 = - \Delta t \frac {\partial{\mathbf{f}^n_y}}{\partial y} + 
 \Delta t \mathbf{F}_{xz}^n = \Delta t \, \mathbf{R}^n, \label{eq:PBW}
\end{equation}
where $\Delta \mathbf{w}^{n} = \mathbf{w}^{n+1}-\mathbf{w}^n$, and where
terms containing transverse flux derivatives and viscous terms are lumped together into 
$\mathbf{F}_{xz}$.
It is important to note that, because of the special structure of the acoustic flux Jacobian,
the inversion of Eqn.~\eqref{eq:PBW} is much simpler than for the
standard Beam-Warming algorithm, which relies on linearization of the full convective flux. 
Component-wise, Eqn.~\eqref{eq:PBW} reads
\begin{subequations}
\begin{numcases}{}
\Delta w^n_1 + \Delta t \frac{\partial}{\partial y} \Delta w^n_3 = \Delta t R_1^n \\
\Delta w^n_2                                          = \Delta t R_2^n \\
\Delta w^n_3 + \Delta t \frac{\partial}{\partial y} ({A_y^a}^n_{31} \Delta w^n_1) + \Delta t \frac{\partial}{\partial y} ({A_y^a}^n_{35} \Delta w^n_5) = \Delta t R_3^n \\
\Delta w^n_4                                          = \Delta t R_4^n \\
\Delta w^n_5                                          = \Delta t R_5^n .
\end{numcases} 
\end{subequations}
Hence, the time increments of entropy and of the transverse velocity components can be 
evaluated explicitly, thus effectively reducing the system of equations to be solved to
\begin{subequations}
\begin{numcases}{}
\Delta w^n_1 + \Delta t \frac{\partial}{\partial y} \Delta w^n_3 = \Delta t R_1^n \label{eq:2by2a} \\
\Delta w^n_3 + \Delta t \frac{\partial}{\partial y} ({A_y^a}^n_{31} \Delta w^n_1) = \Delta t R_3^n - \Delta t \frac{\partial}{\partial y} ({A_y^a}^n_{35} \Delta w^n_5) =: \Delta t \widehat{R}_3^n ,
\end{numcases}
\label{eq:2by2}
\end{subequations}
which, upon discretization of the space derivative operators, yields a $2 \times 2$
block-banded system of equations, whose solution returns
the time increments of $\rho$ and $\rho v$.
Equation~\eqref{eq:2by2} can be further rearranged by formally solving for $\Delta w^n_1$ in~\eqref{eq:2by2a}, to obtain
\begin{equation}
\left(1 - \Delta t^2 {A_y^a}^n_{31} \frac {\partial^2}{\partial y^2} - \Delta t^2 \frac {\partial {A_y^a}^n_{31}}{\partial y} \frac {\partial}{\partial y} \right) \Delta w^n_3 = \Delta t \widehat{R}_3^n - \Delta t^2 \frac {\partial}{\partial y} \left( {A_y^a}^n_{31} R_1^n , \right), \label{eq:1by1} 
\end{equation}
whose solution requires the inversion of a single ordinary banded system of equations, with bandwidth depending on the
accuracy in the approximation of the first and second space derivative operators.
Back substitution into~\eqref{eq:2by2a} then returns the time increment of density.
Although apparently cumbersome, we find the latter formulation to be more computationally efficient 
than the solution of the $2 \times 2$ block system given by Eqn.~\eqref{eq:2by2}, 
while the accuracy is nearly identical. Hence,
Eqn.~\eqref{eq:1by1} is used in all the forthcoming numerical applications.

\subsection{Implicit treatment of viscous terms} \label{sec:viscous}

If needed, viscous terms can also be handled implicitly, using approximate factorization.
For that purpose, we split the viscous flux derivatives in Eqn.~\eqref{eq:NS} into a Laplacian term and a difference thereof
\begin{equation}
\frac{\partial \mathbf{f}_y^v}{\partial y} = \boldsymbol \mu \frac{\partial^2 {\mathbf{v}}}{\partial y^2} + {\boldsymbol{\varphi}_y^v}, \label{eq:vsplit}
\end{equation}
where $\mathbf{v}$ is the vector of primitive variables, $\mathbf{v}=\left[\rho, u, v, w, T\right]$, 
and $\boldsymbol \mu$ is the viscosity matrix,
\begin{equation}
 \boldsymbol{\mu}=
 \begin{bmatrix}
  0& 0& 0& 0& 0&\\              
  0& \mu& 0& 0& 0&\\              
  0& 0& \mu& 0& 0&\\              
  0& 0& 0& \mu& 0&\\              
  0& 0& 0&   0& \frac{\mu Cp}{\Pran T} &\\              
 \end{bmatrix} .
\end{equation}
Freezing for simplicity the viscosity matrix at time step $n$, the following
linearization is considered,
\begin{equation}
\left( \boldsymbol \mu \frac{\partial^2 {\mathbf{v}}}{\partial y^2} \right)^{n+1} \approx
\left( \boldsymbol \mu \frac{\partial^2 {\mathbf{v}}}{\partial y^2} \right)^n + 
\boldsymbol \mu^n \frac{\partial^2 {\mathbf{P} \Delta \mathbf{w}^n}}{\partial y^2} ,
\end{equation}
where $\mathbf{P}$ is the Jacobian of the conservative-to-primitive variables transformation
\begin{equation}
\mathbf{P}=\frac{\partial \mathbf{v}}{\partial\mathbf{w}}= 
 \begin{bmatrix}
  1& 0& 0& 0& 0&\\              
  -\frac{u}{\rho}& \frac{1}{\rho}& 0& 0& 0&\\              
  -\frac{v}{\rho}&0& \frac{1}{\rho}& 0& 0&\\              
  -\frac{w}{\rho}&0&0&\frac{1}{\rho}& 0&\\              
  \frac{-T s}{\rho c_v}&
  0&0&0& \frac{T}{\rho c_v}&\\              
 \end{bmatrix} . \label{eq:P}
\end{equation}
Following similar steps as done to arrive at Eqn.~\eqref{eq:PBW}, the previous linearization yields
\begin{equation}
\left( \mathbf{I} + \Delta t \frac{\partial}{\partial y} {\mathbf{A}_y^a}^n - \Delta t \, \boldsymbol{\mu}^n \frac{\partial^2}{\partial y^2} \mathbf{P}^{n} \right) \Delta \mathbf{w}^{n} = \Delta t \, \mathbf{R}^n, \label{eq:PBWV}
\end{equation}
which can be approximately factorized as follows
\begin{equation}
 \left(\mathbf{I}+\Delta t \frac{\partial}{\partial y} {\mathbf{A}_y^a}^n \right)
 \left(\mathbf{I}
 -\Delta t \, \boldsymbol{\mu}^n \frac{\partial^2}{\partial y^2} \mathbf{P}^{n} \right) \Delta \mathbf{w}^n = 
 \Delta t \, \mathbf{R}^n. \label{eq:PBWVfac}
\end{equation}
Inversion of Eqn.~\eqref{eq:PBWVfac} can be then carried out into two sequential sub-steps,
\begin{eqnarray}
  \left(\mathbf{I}+\Delta t \frac{\partial}{\partial y} {\mathbf{A}_y^a}^n \right) \widetilde{\Delta \mathbf{w}}^n &=& \Delta t \mathbf{R}^n, \label{eq:PBWVfac2a} \\
  \left(\mathbf{I}
 -\Delta t \, \boldsymbol{\mu}^n \frac{\partial^2}{\partial y^2} \mathbf{P}^{n} \right) {\Delta \mathbf{w}^n} &=& 
 \widetilde{\Delta \mathbf{w}}^n , \label{eq:PBWVfac2b}
\label{eq:PBWVfac2}
\end{eqnarray}
whereby the provisional time increment $\widetilde{\Delta \mathbf{w}}^n$ is first evaluated 
through the inversion procedure for the convective fluxes described in section~\ref{sec:beam}.
The actual time increment $\Delta \mathbf{w}^n$ is then evaluated by inverting the viscous implicit operator at the 
left-hand-side of Eqn.~\eqref{eq:PBWVfac2b} which, in light of the special structure
of the Jacobian matrix given in Eqn.~\eqref{eq:P}, can be carried out sequentially, as follows
\begin{subequations}
 \begin{numcases}{}
  \Delta w_1^n = \widetilde{\Delta w_1}^n \\
  \left(1-\mu^n_{22} \Delta t \frac{\partial^2}{\partial y^2} P_{22}^n \right) \Delta w_2^n = \widetilde{\Delta w_2}^n + \mu_{22}^n \Delta t  \frac{\partial^2}{\partial y^2} \left( P_{21}^n \Delta w_1^n \right) \\
  \left(1-\mu^n_{33} \Delta t \frac{\partial^2}{\partial y^2} P_{33}^n \right) \Delta w_3^n = \widetilde{\Delta w_3}^n + \mu_{33}^n \Delta t  \frac{\partial^2}{\partial y^2} \left( P_{31}^n \Delta w_1^n \right) \\
  \left(1-\mu^n_{44} \Delta t \frac{\partial^2}{\partial y^2} P_{44}^n \right) \Delta w_4^n = \widetilde{\Delta w_4}^n + \mu_{44}^n \Delta t  \frac{\partial^2}{\partial y^2} \left( P_{41}^n \Delta w_1^n \right) \\
  \left(1-\mu^n_{55} \Delta t \frac{\partial^2}{\partial y^2} P_{55}^n \right) \Delta w_5^n = \widetilde{\Delta w_5}^n + \mu_{55}^n \Delta t  \frac{\partial^2}{\partial y^2} \left( P_{51}^n \Delta w_1^n \right) 
 \end{numcases} 
\label{eq:avti}
\end{subequations}
The inversion of four standard narrow-banded systems of equations is thus required for the purpose.
We point out that the present procedure is again different than the original Beam-Warming procedure,
which relies on linearization of the full viscous flux vectors, hence requiring the
inversion of block-banded systems. However, we have found that numerical robustness is very weakly affected
by the approximations herein made.

\subsection{Multiple space dimensions} \label{sec:multid}

As done for the case of a single space dimension, the acoustic and viscous time limitations can be 
removed in more than one direction through direction-wise factorization of the implicit operators. 
For instance, assuming that all space directions are handled in semi-implicit fashion, Eqn.~\eqref{eq:PBWVfac} is replaced by
\begin{equation}
\mathbf{L}^n \Delta \mathbf{w}^n = \mathbf{R}^n, \label{eq:multid}
\end{equation}
where
\begin{eqnarray}
 \mathbf{L}^n = 
 && \left(\mathbf{I}+\Delta t \frac{\partial}{\partial x} {\mathbf{A}_x^a}^n \right)
   \left(\mathbf{I}+\Delta t \frac{\partial}{\partial y} {\mathbf{A}_y^a}^n \right)
   \left(\mathbf{I}+\Delta t \frac{\partial}{\partial z} {\mathbf{A}_z^a}^n \right) \cdot \nonumber \\
 && \left(\mathbf{I} -\Delta t \boldsymbol{\mu}^n \frac{\partial^2}{\partial x^2} \mathbf{P}^{n} \right) 
   \left(\mathbf{I} -\Delta t \boldsymbol{\mu}^n \frac{\partial^2}{\partial y^2} \mathbf{P}^{n} \right) 
   \left(\mathbf{I} -\Delta t \boldsymbol{\mu}^n \frac{\partial^2}{\partial z^2} \mathbf{P}^{n} \right) . 
   \label{eq:LHS}
\end{eqnarray}
Hence, repeated application of the procedures developed in the previous two sections is sufficient.
Practical application of Eqn.~\eqref{eq:LHS} requires some caution, as the order in which the 
various inversions are carried out is not immaterial. We have found that, in order to remove 
possible spurious anisotropies, it is a good practice to shuffle the order of the implicit 
left-hand-side operators. 

\subsection{Time integration} \label{sec:rk}

Time accuracy and stability enhancement is typically obtained by Runge-Kutta schemes as wrapper to 
one-step implicit procedures outlined in the previous paragraphs.
Low-storage algorithms are a popular choice, and here we consider for example Wray's three-stage, third-order scheme~\citep{orlandi_00}, adapted to semi-implicit integration of the convective terms,
\begin{equation}
 \mathbf{L}^{(\ell)} \Delta \mathbf{w}^{(\ell)} = \alpha_{\ell} \Delta t \mathbf{R}^{(\ell-1)} + \beta_{\ell} \Delta t \mathbf{R}^{(\ell)}, \quad \ell=0,1,2,
\label{eq:RK}
\end{equation}
where $\Delta \mathbf{w}^{(\ell)} = \mathbf{w}^{(\ell+1)} -  \mathbf{w}^{(\ell)}$, $\mathbf{w}^{(0)}=\mathbf{w}^{n}$,
$\mathbf{w}^{n+1}=\mathbf{w}^{(3)}$, the left-hand-side implicit operator is a generalization of Eqn.~\eqref{eq:LHS}, namely
\begin{eqnarray}
 \mathbf{L}^{(\ell)} = 
 && \left(\mathbf{I}+\gamma_{\ell} \Delta t \frac{\partial}{\partial x} {\mathbf{A}_x^a}^{(\ell)} \right)
   \left(\mathbf{I}+\gamma_{\ell} \Delta t \frac{\partial}{\partial y} {\mathbf{A}_y^a}^{(\ell)} \right)
   \left(\mathbf{I}+\gamma_{\ell} \Delta t \frac{\partial}{\partial z} {\mathbf{A}_z^a}^{(\ell)} \right) \cdot \nonumber \\
 && \left(\mathbf{I} -\gamma_{\ell} \Delta t \boldsymbol{\mu}^{(\ell)} \frac{\partial^2}{\partial x^2} \mathbf{P}^{(\ell)} \right) 
   \left(\mathbf{I} -\gamma_{\ell} \Delta t \boldsymbol{\mu}^{(\ell)} \frac{\partial^2}{\partial y^2} \mathbf{P}^{(\ell)} \right) 
   \left(\mathbf{I} -\gamma_{\ell} \Delta t \boldsymbol{\mu}^{(\ell)} \frac{\partial^2}{\partial z^2} \mathbf{P}^{(\ell)} \right) , \nonumber
   \label{eq:LHSl}
\end{eqnarray}
and the integration coefficient are $\alpha_{\ell} = (0, 17/60,-5/12)$, $\beta_{\ell} = (8/15, 5/12, 3/4)$,
$\gamma_{\ell} = \alpha_{\ell} + \beta_{\ell}$. 
We have found this time stepping scheme to work well in practice, 
however because of the 
partial flux linearization, the method is only formally first-order accurate in time.

A genuinely third-order accurate semi-implicit Runge-Kutta scheme was derived by \citet{nikitin_06},
which can be conveniently cast as follows
\begin{subequations}
\begin{numcases}{}
 \mathbf{L}^{n} \Delta \mathbf{w}^{(1)} = \frac 23 \Delta t \mathbf{R}^{n} \\
 \mathbf{L}^{(1)} \Delta \mathbf{w}^{(2)} = - \left( \mathbf{w}^{(1)} - \mathbf{w}^{n}\right) + \frac 13 \Delta t \mathbf{R}^{n} + \frac 13 \Delta t \mathbf{R}^{(1)} \\
 \Delta \mathbf{w}^{(3)} = \frac 12 \left( \mathbf{w}^{(2)} - \mathbf{w}^{n} \right) - \frac 32 \alpha \Delta \mathbf{w}^{(2)} \\
 \mathbf{L}^{(3)} \Delta \mathbf{w}^{(4)} = - \left( \mathbf{w}^{(3)} - \mathbf{w}^{n}\right) + \frac 14 \Delta t \mathbf{R}^{n} + \frac 34 \Delta t \mathbf{R}^{(1)} \\
 \mathbf{L}^{(4)} \Delta \mathbf{w}^{(5)} = - \left( \mathbf{w}^{(4)} - \mathbf{w}^{n}\right) + \frac 14 \Delta t \mathbf{R}^{n} + \frac 34 \Delta t \mathbf{R}^{(2)} ,
\end{numcases} 
\label{eq:Nikitin}
\end{subequations}
where $\gamma_{\ell} = \gamma$ is the same for all sub-steps, and $\alpha$ are free parameters (hereafter, we assume $\alpha=1$, $\gamma=0.6$).
With respect to Wray's algorithm, Eqn.~\eqref{eq:Nikitin} is not in low-storage form (although it can be implemented using three arrays only), and it involves an additional inversion, but no additional evaluation of the explicit operator. Despite the slight computational overhead, 
all the following analysis and numerical experiments are carried out with algorithm~\eqref{eq:Nikitin} because of its higher formal accuracy.

\subsection{Stability analysis} \label{sec:stab}

The stability of the semi-implicit algorithm herein developed is
here analyzed within the simplified setting of the linearized inviscid acoustic equations in the presence 
of a mean flow $u_0$, which can be cast as
\begin{equation}
\frac {\partial {\bf v}}{\partial t} + {\bf A} 
\frac {\partial {\bf v}}{\partial x} = 0, \quad
{\bf v} = 
\begin{bmatrix}
\rho' \\ u'
\end{bmatrix}, \quad
{\bf A} = 
\begin{bmatrix}
u_0 & \rho_0 \\ c_0^2/\rho_0 & u_0
\end{bmatrix} , \label{eq:acoustics}
\end{equation}
where the subscript $0$ refers to the unperturbed state, and primes to fluctuations thereof.
A semi-implicit discretization of \eqref{eq:acoustics} can be obtained by considering the 
linearized counterpart of the partial flux Jacobian~\eqref{eq:acoustic_jac}, namely
\begin{equation}
{\bf A}^a = 
\begin{bmatrix}
u_0 & \rho_0 \\ c_0^2/\rho_0 & 0
\end{bmatrix} . \label{eq:pjac}
\end{equation}
Backward Euler discretization of Eqn.~\eqref{eq:acoustics} then yields
\begin{equation}
\left( {\bf I} - \Delta t {\bf A}^a \frac{\partial}{\partial x} \right) \Delta {\bf v}^n = - \Delta t {\bf A} \frac{\partial {\bf v}^n}{\partial x}. \label{eq:BE}
\end{equation}
Transforming Eqn.~\eqref{eq:BE} to Fourier space with
the token ${\bf v} (x,t) = \hat{{\bf v}}(t) e^{i k x}$ yields the amplification matrix of the scheme 
\begin{equation}
{\bf G} = {\bf I} - \left( {\bf I} - i {\Delta t} \tilde{k} {\bf A}^a \right)^{-1} i {\Delta t} \tilde{k} {\bf A}, 
\end{equation}
where ${\bf v}^{n+1} = {\bf G} {\bf v}^n$, and $\tilde{k}$ is the modified wavenumber 
corresponding to the discretization of the space first derivative operator~\citep{vichnevetsky_82}.
Von Neumann's stability condition requires that both eigenvalues of ${\bf G}$ are
no larger than unity in modulus. Assuming for instance second-order central 
differencing (i.e. $\tilde{k} h = \sin (k h)$), it turns out that the scheme~\eqref{eq:BE} is unconditionally
stable for $M_0 = u_0/c_0 \lesssim 1$.
A similar analysis can be carried out (details are omitted) for the Runge-Kutta time stepping scheme
of Eqn.~\eqref{eq:Nikitin}. In the case of explicit time integration (i.e. $\gamma=0$) the
scheme is stable for $\mathrm{CFL} \lesssim \sqrt{3}$, where $\mathrm{CFL} = (u_0+c_0) \Delta t / h$. In the case of semi-implicit time integration
(with $\gamma=0.6$, $\alpha=1$) unconditional stability is achieved for $M_0 \lesssim 0.525$.

\begin{figure}
 \centering
 \psfrag{x}[t][][1.0]{$k h$}
 \psfrag{y}[b][][1.0]{$g_1$}
 (a)
 \includegraphics[width=3.4cm,clip,angle=270]{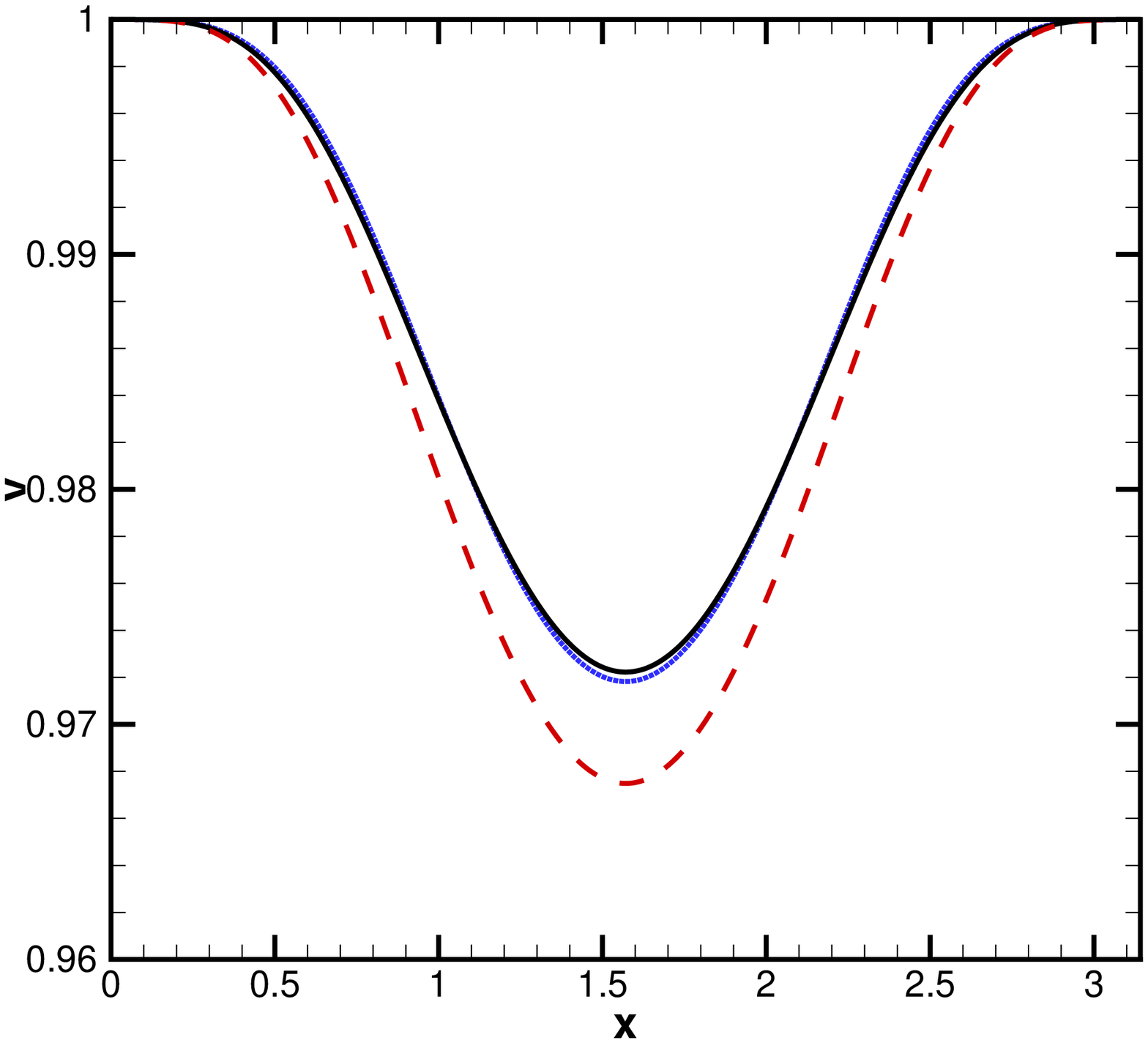}
 (b)
 \includegraphics[width=3.4cm,clip,angle=270]{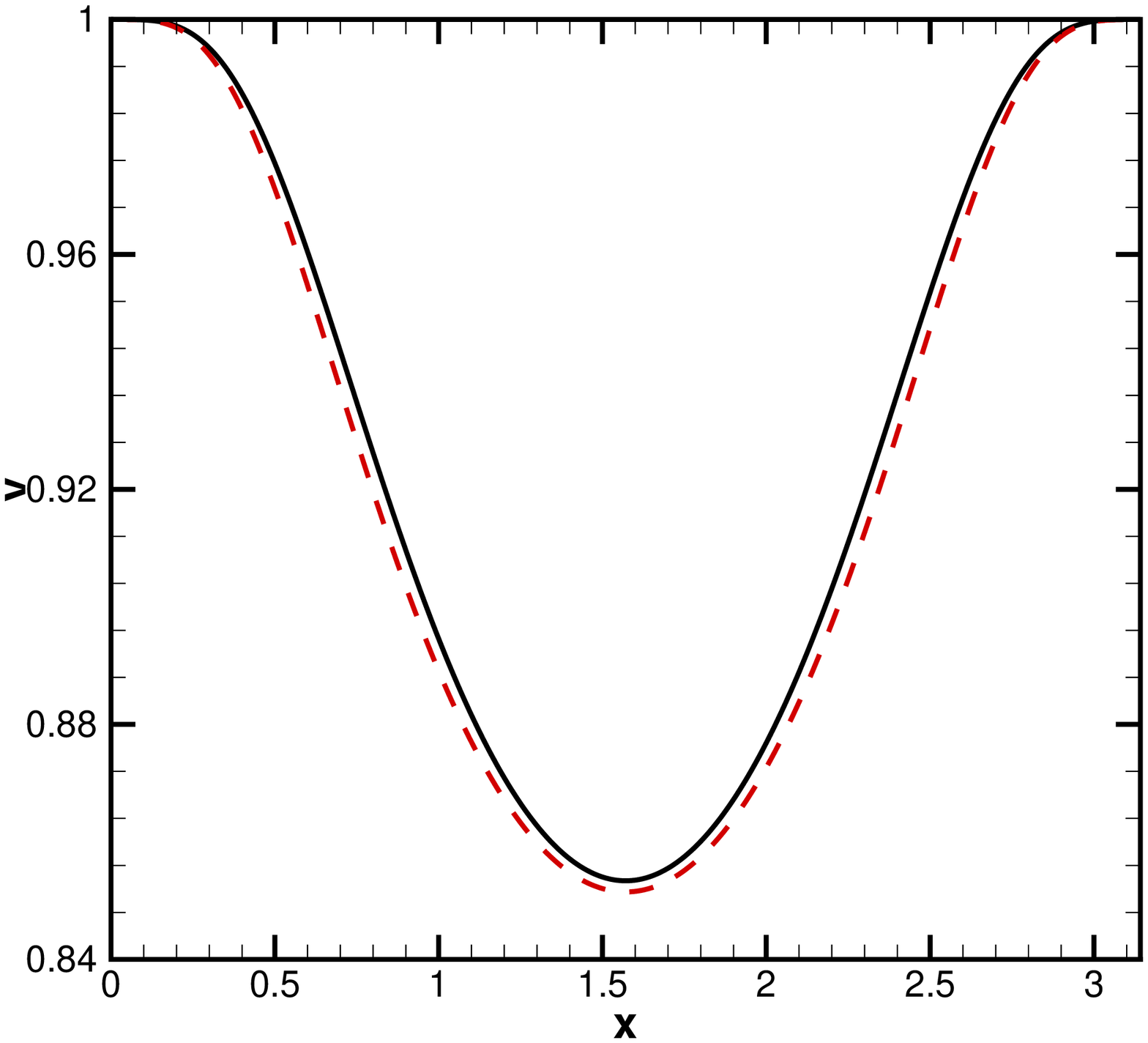}
 (c)
 \includegraphics[width=3.4cm,clip,angle=270]{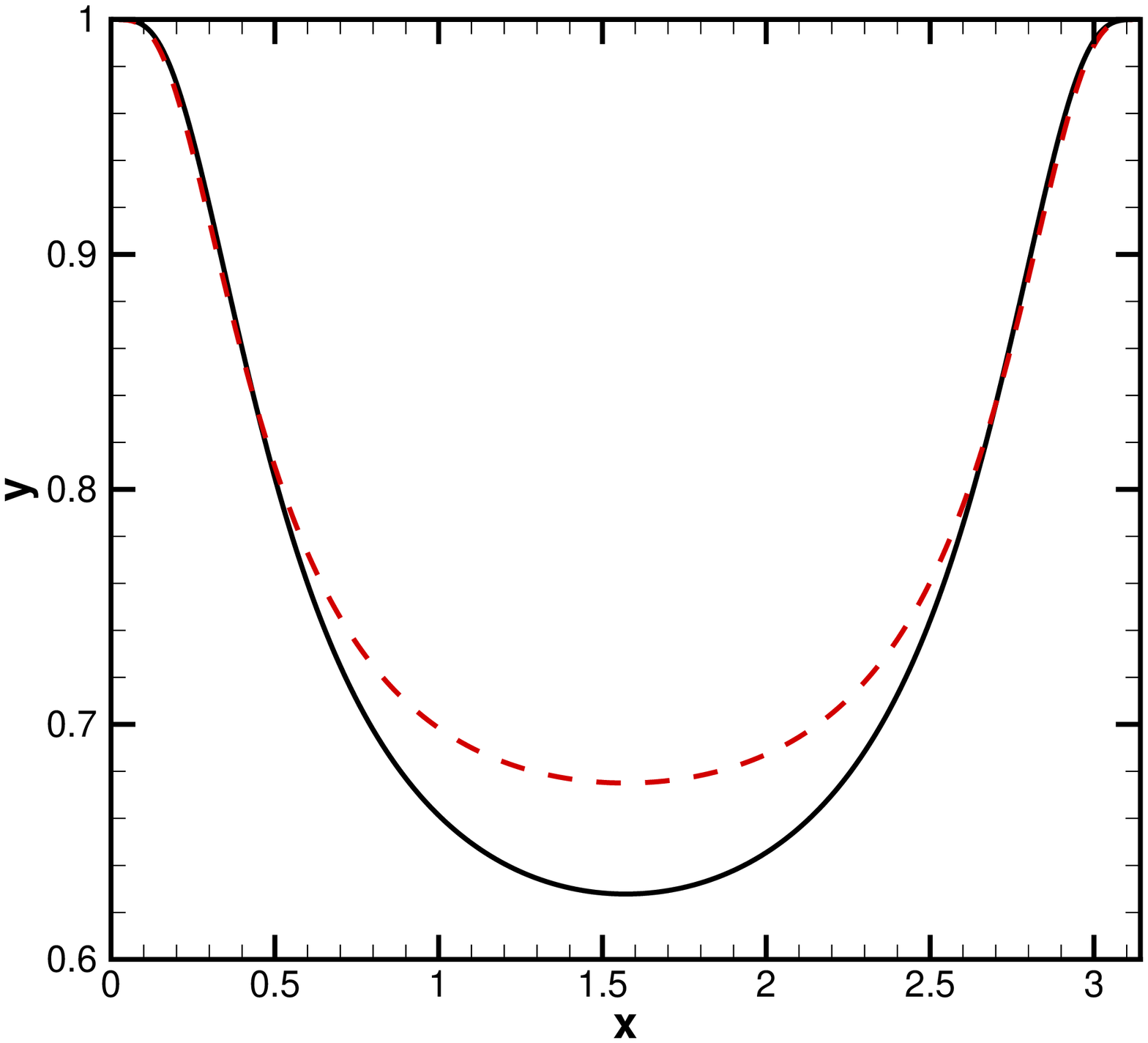} \\
 \caption{Smallest eigenvalue of amplification matrix at $\mathrm{CFL}=1$ (a), $\mathrm{CFL}=2$ (b), $\mathrm{CFL}=5$ (c), 
  for explicit Runge-Kutta time integration (dotted lines), semi-implicit time integration (with $\alpha=1$, $\gamma=0.6$, solid lines),
  and fully implicit Beam-Warming scheme (dashed lines), at Mach number $M_0=0.3$. Curves are only shown for stable schemes.}
 \label{fig:gfac}
\end{figure}

To provide an idea of the accuracy of the algorithm, in Fig.~\ref{fig:gfac} 
we show the smallest eigenvalues of the amplification matrix at various Courant numbers
for explicit and semi-implicit Runge-Kutta time integration. For reference, the amplification factor
of the baseline Beam-Warming algorithm is also shown.
At CFL numbers lower than the stability limit for explicit discretization (panel (a)), the 
semi-implicit and the fully explicit algorithms have similar performance, whereas the Beam-Warming algorithm has
somewhat higher diffusion. At higher Courant numbers the explicit scheme goes unstable, and
semi-implicit and fully implicit scheme have similar performance, with slightly less diffusive behavior
of Beam-Warming at higher $\mathrm{CFL}$. Notably, all schemes have unit amplification factor 
at the Niquist limit ($k h = \pi$), hence they are not dissipative in the sense of Kreiss.
This is the reason why schemes of the Beam-Warming family are typically used with explicit addition of
artificial diffusion terms~\citep{beam_76,hirsch_07}. 

\subsection{Spatial discretization}

All the convective derivatives at the right-hand-side operator defined in Eqn.~\eqref{eq:NS} are discretized
using conservative, energy-preserving formulas~\citep{pirozzoli_10},
based on application of standard central difference approximations to
the fully expanded form of the convective derivatives~\citep{kennedy_08}.
In the explicit case this discretization allows to exactly preserve
the total kinetic energy from convection, and conserve the entropy variance 
in the inviscid limit, hence providing strong nonlinear stability to the algorithm without introducing 
any numerical diffusion~\citep{honein_04,pirozzoli_11a}.
We have found that this feature is very important to prevent nonlinear divergence 
caused by accumulation of aliasing errors,
especially in light of the fact that the semi-implicit algorithms herein dealt with 
have zero numerical diffusion at the highest resolved wavenumbers.
Hence, no explicit addition of artificial diffusion is needed for the semi-implicit algorithm herein developed.
Viscous terms are also expanded to Laplacian form and discretized by means of central formulas~\citep{lele_92}.

Consistency requires that the same finite-difference operators are applied to the implicit
and the implicit operators. Hence, for the sake of simplicity in the present work we only consider 
second-order space discretizations, which only require the inversion of standard tridiagonal matrices.
However, extension of the algorithm to higher-order spatial accuracy is straightforward, and it 
can be achieved by considering compact-difference approximations with narrow stencil~\citep{hirsch_07}, 
or by simply widening the stencil.
In the latter case, fourth-order order spatial accuracy can be achieved at the price of
inverting standard pentadiagonal matrices, and so on.

\subsection{Computational efficiency}

\begin{table}
 \centering
 \begin{tabular}{lc}
 \hline
 Scheme & CPU/CPU$_{\mathrm{EXPL}}$ \\
 \hline
EXPL           &     1.    \\
ATI            &    1.14   \\
ATI-CYC        &    1.16   \\
AVTI           &    1.32   \\
AVTI-CYC       &    1.37   \\
BW             &    1.67   \\
BW-CYC         &    2.21   \\
BWV            &    1.87   \\
BWV-CYC        &    2.33   \\
 \hline
 \end{tabular}
 \caption{Computational cost for implicit schemes compared to fully explicit discretization. Figures refer to implicit treatment of a single space direction.}
\label{tab:cost}
\end{table}

Achieving higher computational efficiency is obviously the main motivation for 
using implicit algorithms, which are inherently more computationally intensive than explicit ones. 
Computational cost figures for the present semi-implicit algorithm and for the
Beam-Warming scheme are listed in table~\ref{tab:cost}, as a fraction of 
the cost for the baseline explicit algorithm. Cost estimates are given 
for implicit treatment of convective terms only, 
and for simultaneous treatment of convective and viscous terms, 
referring to a single space direction. 
Also for ease of later reference, we use the following notation to distinguish the various schemes.
The semi-implicit scheme herein developed is referred to as either ATI (acoustic terms-implicit, as in Eqn.~\eqref{eq:PBW}), 
or ATVI in the case that both convective and viscous terms are handled implicitly (Eqn.~\eqref{eq:PBWV}).
As a basis of comparison, cost figures for the Beam-Warming (BW) scheme, also
with implicit treatment of the viscous terms (BWV) are reported.
Cost figures are provided for both the case of 
periodic (CYC) and non-periodic boundary conditions.
It should be noted that the cost estimates refer to actual parallel computations, 
and also include the computational overhead for data transposition across 
processors in non-contiguous space directions. Of course, precise figures may change
depending on the specific implementation of the algorithm and/or 
machine architecture, but we trust that the numbers listed in the table provide a
reasonably robust estimate. It appears that the computational overhead 
of the ATI algorithm is rather limited, hence implicit 
treatment of a given space direction is computationally advantageous provided the 
attainable time step is at least $20\%$ higher than for fully explicit. 
Substantial improvement of computational efficiency over standard Beam-Warming 
discretization is also apparent, for comparable expected accuracy (recalling Fig.~\ref{fig:gfac}).

\section{Numerical results} \label{sec:results}

The performance of the semi-implicit algorithm herein developed is tested 
through application to a series of canonical compressible turbulent flows, 
in order of increasing physical complexity.

\subsection{Isotropic turbulence} \label{sec:ict}

\begin{figure}
 \centering
 \psfrag{X}[t][][1.0]{$t/\tau$}
 \psfrag{Y}[b][][1.0]{$K/K_0$}
 (a)
 \includegraphics[width=5cm,clip,angle=270]{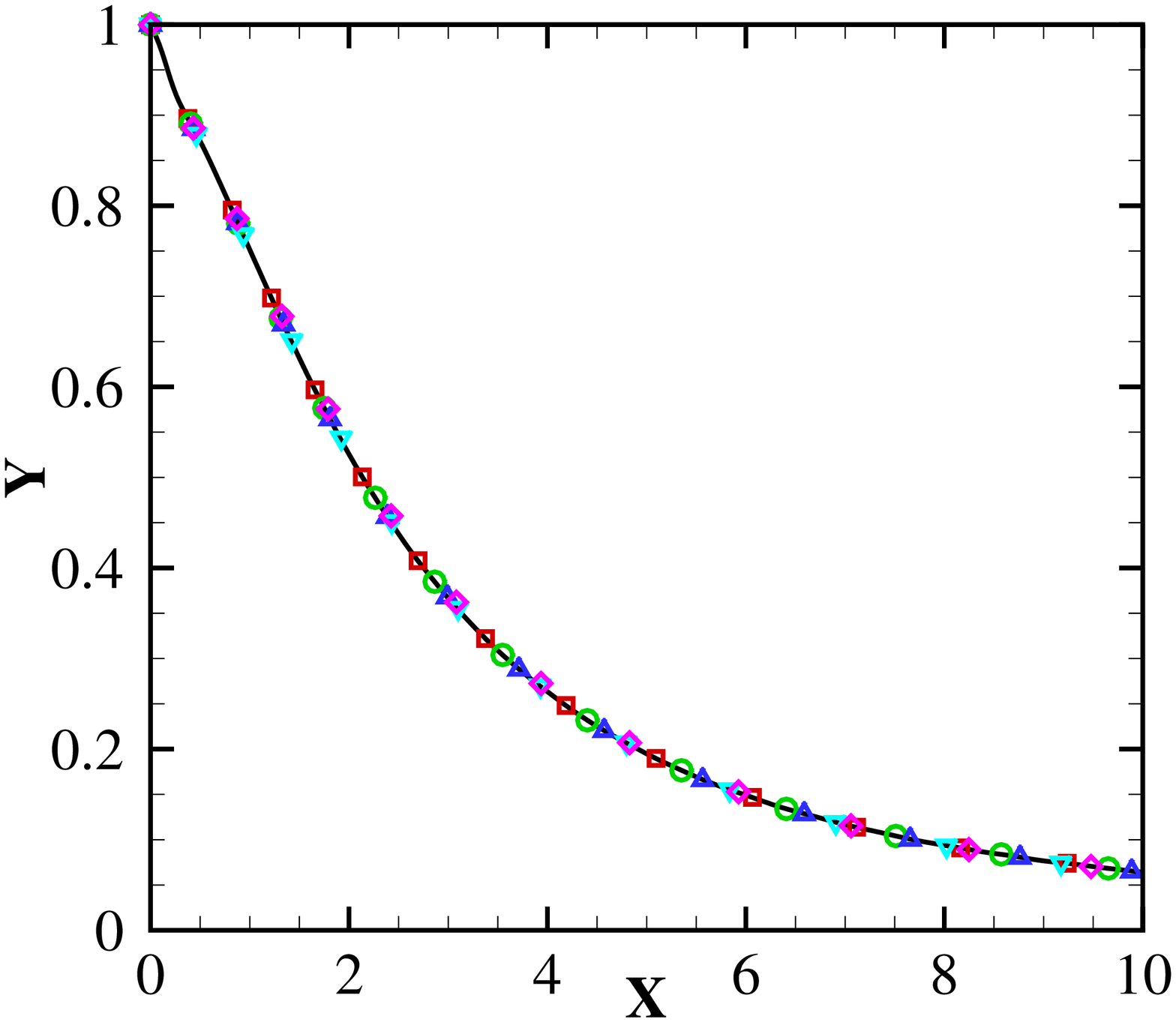}
 \psfrag{X}[t][][1.0]{$t/\tau$}
 \psfrag{Y}[b][][1.0]{$p_{\mathrm{rms}}/p_0$}
 (b)
 \includegraphics[width=5cm,clip,angle=270]{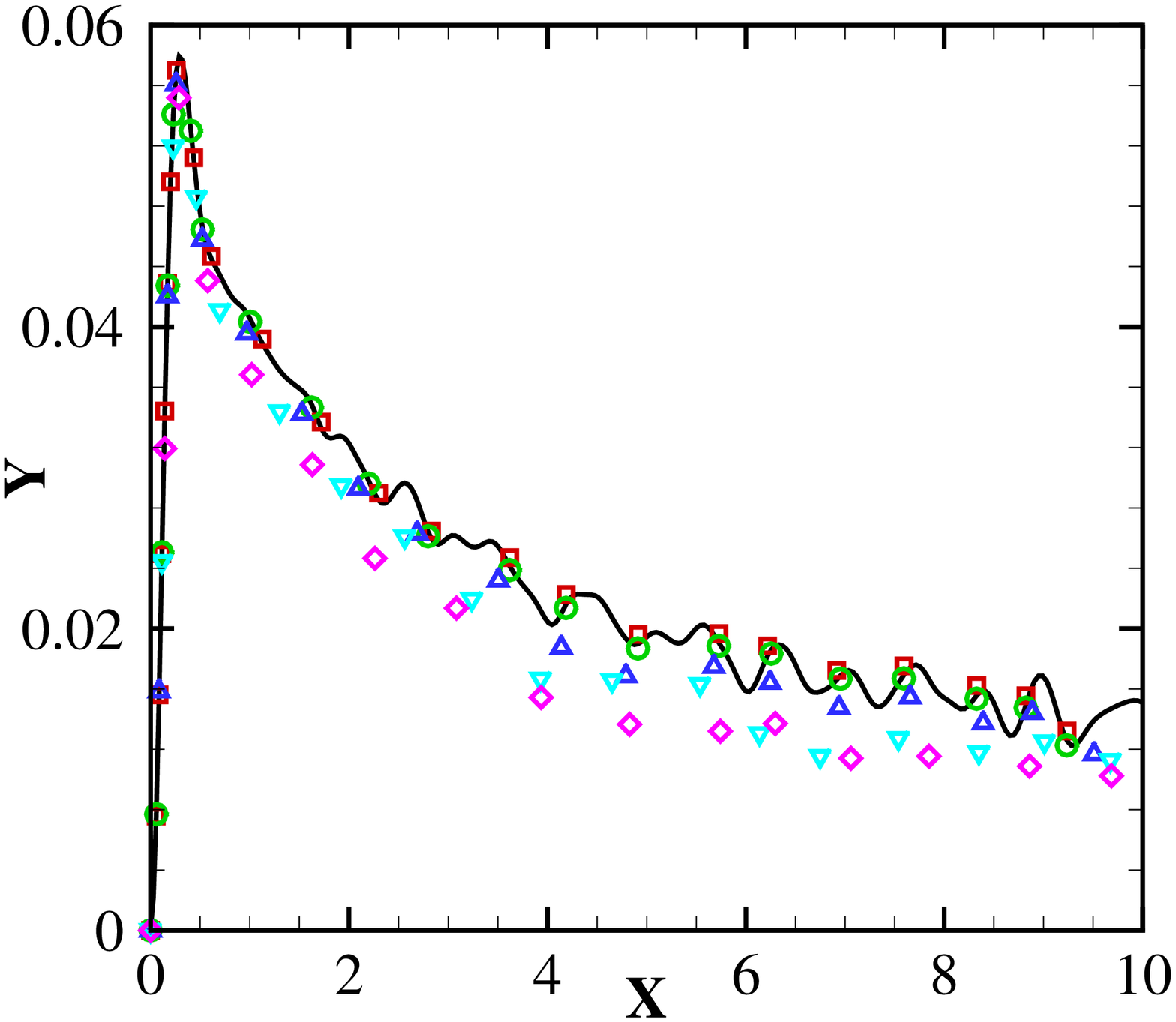}\\
\vskip1.em
 \psfrag{X}[t][][1.0]{$k$}
 \psfrag{Y}[b][][1.0]{$E_u(k)$}
 (c)
 \includegraphics[width=5cm,clip,angle=270]{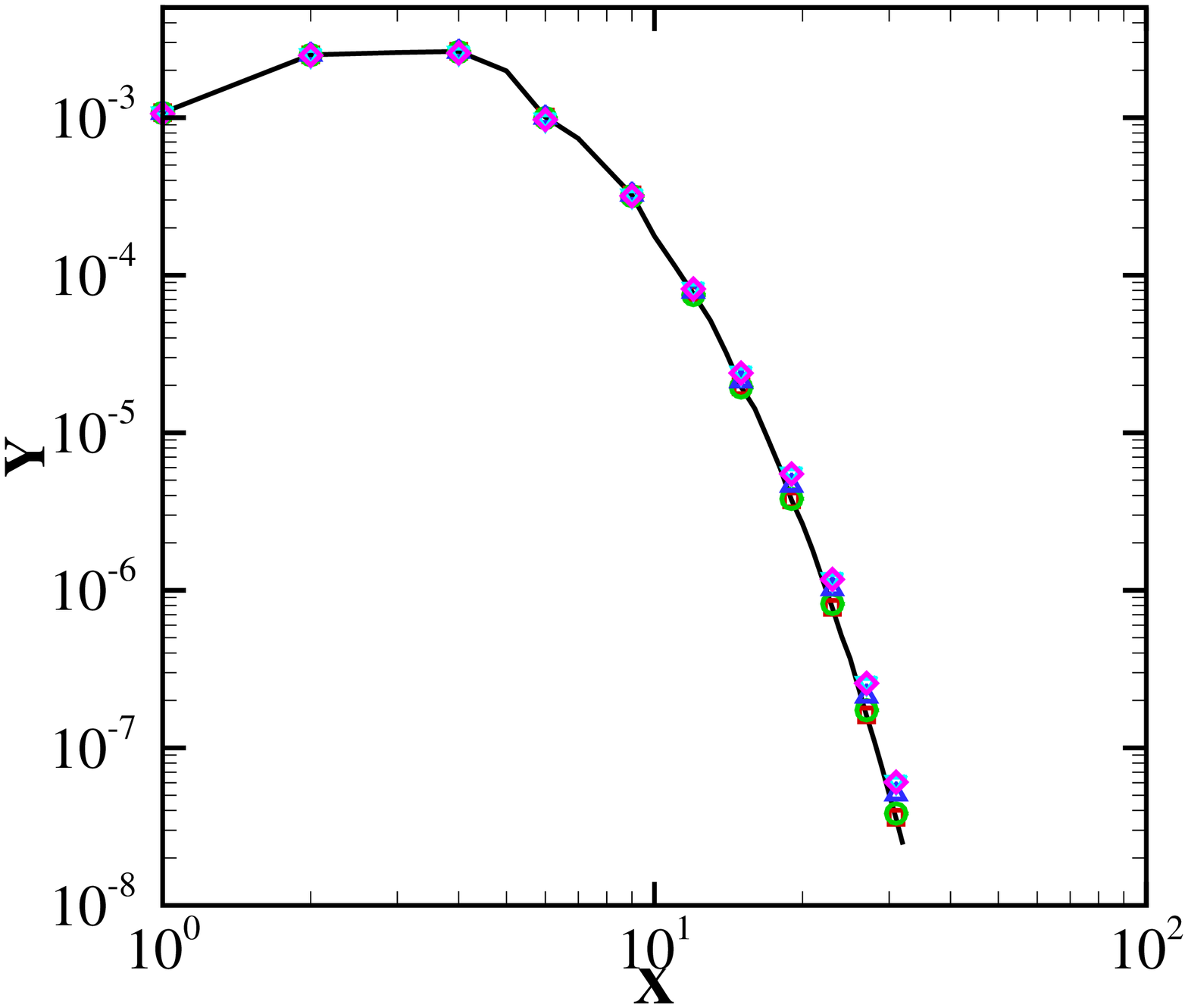}
 \psfrag{X}[t][][1.0]{$k$}
 \psfrag{Y}[b][][1.0]{$E_p(k)$}
 (d)
 \includegraphics[width=5cm,clip,angle=270]{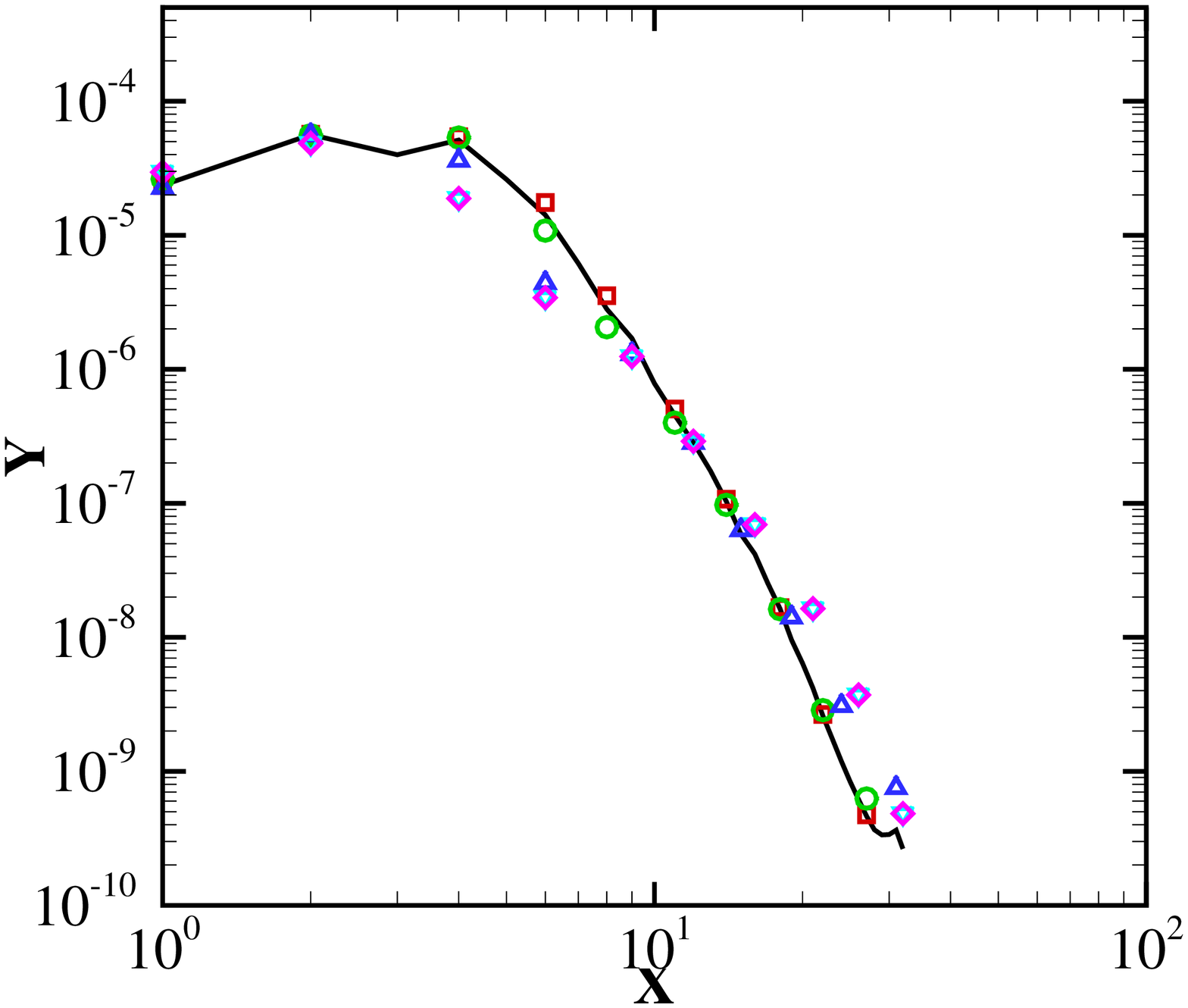}\\
\vskip1.em
 \caption{Numerical simulations of homogeneous isotropic turbulence at $M_t=0.3$, $k_0=4$, $Re_{\lambda}=30$, with ATI-XYZ scheme. Time history of turbulence kinetic energy (a), and pressure variance (b), and spectra of velocity (c) and pressure fluctuations (d) at $t/\tau=5$.
 Solid lines denoted reference results obtained with explicit time discretization at $\mathrm{CFL}=1$.
 Symbols denote results obtained with ATI scheme at 
 $\mathrm{CFL} = 1$ (squares), 
 $\mathrm{CFL} = 2$ (circles), 
 $\mathrm{CFL} = 3$ (triangles), 
 $\mathrm{CFL} = 4$ (down-triangles), 
 $\mathrm{CFL} = 5$ (diamonds).
} 
 \label{fig:ICT_ATI}
\end{figure}
\begin{figure}
 \centering
 \psfrag{X}[t][][1.0]{$t/\tau$}
 \psfrag{Y}[b][][1.0]{$K/K_0$}
 (a)
 \includegraphics[width=5cm,clip,angle=270]{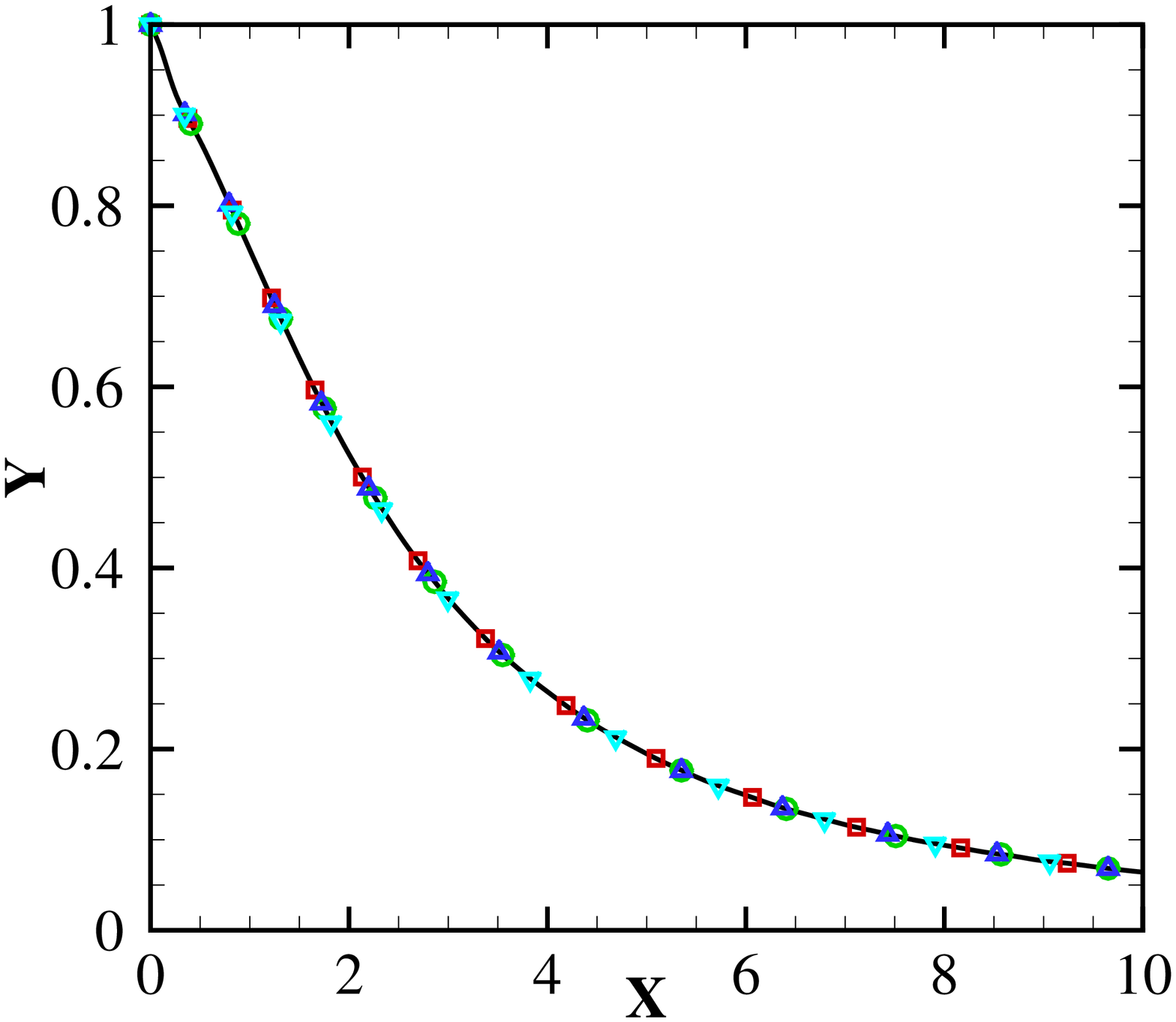}
 \psfrag{X}[t][][1.0]{$t/\tau$}
 \psfrag{Y}[b][][1.0]{$p_{\mathrm{rms}}/p_0$}
 (b)
 \includegraphics[width=5cm,clip,angle=270]{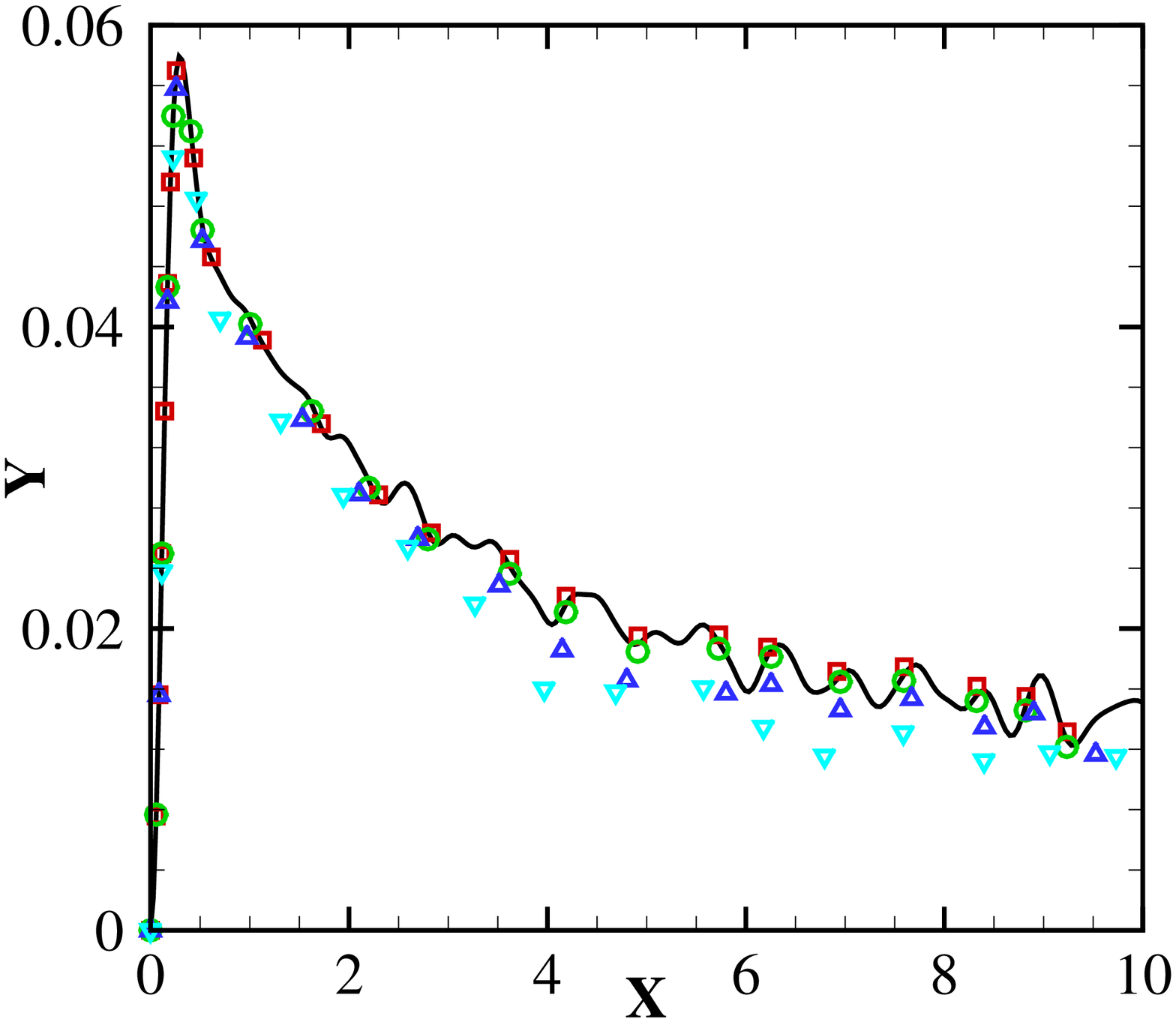}\\
\vskip1.em
 \psfrag{X}[t][][1.0]{$k$}
 \psfrag{Y}[b][][1.0]{$E_u(k)$}
 (c)
 \includegraphics[width=5cm,clip,angle=270]{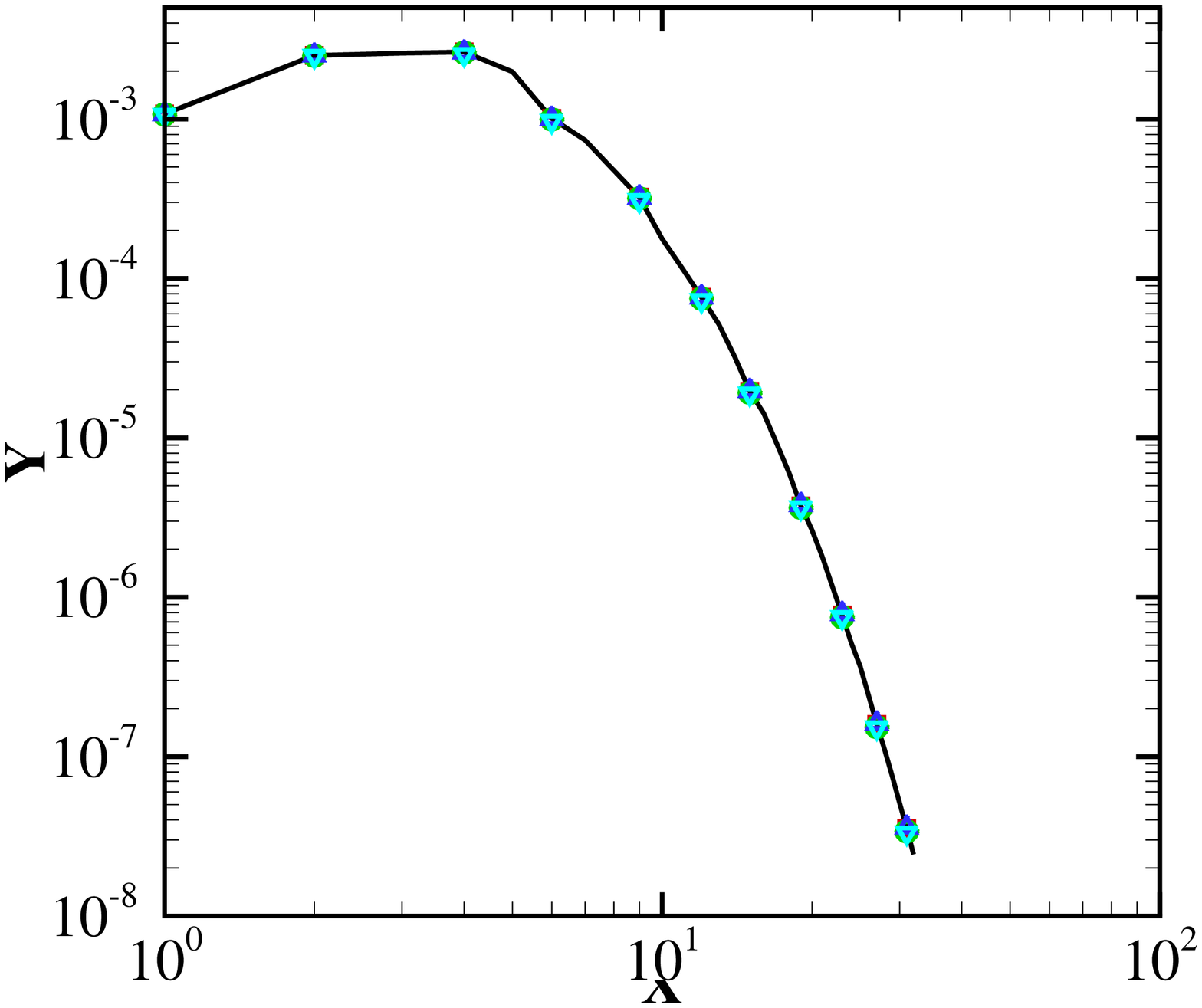}
 \psfrag{X}[t][][1.0]{$k$}
 \psfrag{Y}[b][][1.0]{$E_p(k)$}
 (d)
 \includegraphics[width=5cm,clip,angle=270]{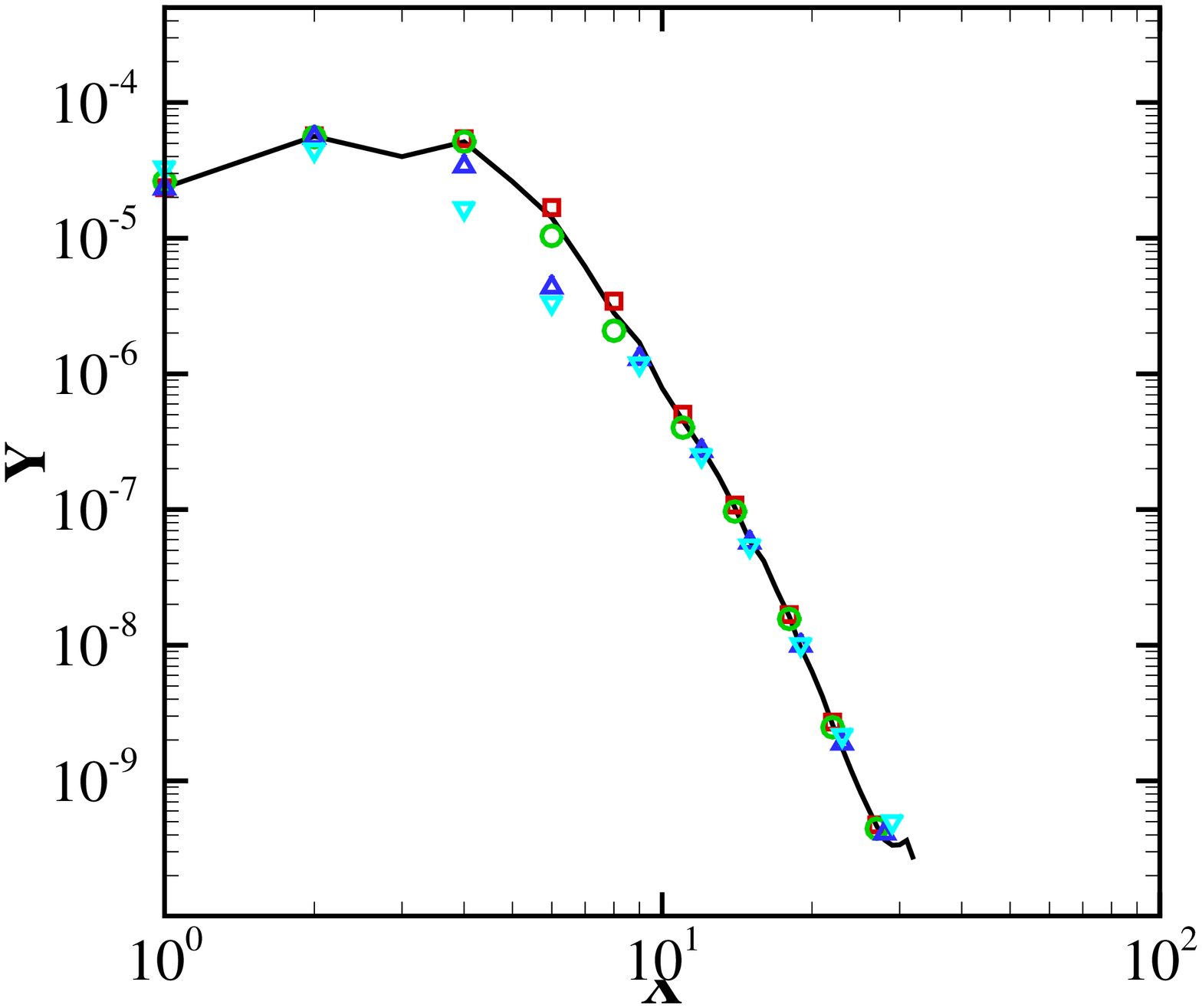}\\
\vskip1.em
 \caption{Numerical simulations of homogeneous isotropic turbulence at $M_t=0.3$, $k_0=4$, $Re_{\lambda}=30$, with BW-XYZ scheme. Time history of turbulence kinetic energy (a), and pressure variance (b), and spectra of velocity (c) and pressure fluctuations (d) at $t/\tau=5$.
 Solid lines denoted reference results obtained with explicit time discretization at $\mathrm{CFL}=1$.
 Symbols denote results obtained with BW scheme at 
 $\mathrm{CFL} = 1$ (squares), 
 $\mathrm{CFL} = 2$ (circles), 
 $\mathrm{CFL} = 3$ (triangles), 
 $\mathrm{CFL} = 4$ (down-triangles), 
} 
 \label{fig:ICT_BW}
\end{figure}
Numerical simulations of homogeneous isotropic turbulence have been frequently
carried out to evaluate the properties of numerical schemes for turbulent flows~\citep{shoeybi_10}.
DNS are here carried out in a triply periodic $(2 \pi)^3$ box,
discretized with $64^2$ collocation points. At the initial time pressure and density are 
taken to be uniform, and solenoidal velocity perturbations are added according to
the procedure introduced by \citet{blaisdell_91}, with prescribed three-dimensional energy spectrum
\begin{equation}
E(k) = 16 \sqrt{\frac 2{\pi}} \frac {u_0^2}{k_0} \left( \frac {k^4}{k_0} \right)^4 e^{-2 (k/k_0)^2},
\end{equation}
where $k_0 = 4$ is the most energetic mode. The initial turbulent Mach number 
is given by $M_{t0} = \sqrt{3} u_0/c_0 = 0.3$, and the Reynolds number based on the Taylor microscale
is $\Rey_{\lambda} = 2 \rho_0 u_0 / (\mu_0 k_0) = 30$. Time is made nondimensional
with respect to the eddy turnover time $\tau = 2 \sqrt{3} / (k_0 M_{t0} c_0)$.

The results obtained with ATI and BW discretization in all space directions are shown in Figs.~\ref{fig:ICT_ATI} and \ref{fig:ICT_BW}, respectively, at various Courant numbers. Stable results are obtained for $\mathrm{CFL} \lesssim 5.1$ for ATI, and $\mathrm{CFL} \lesssim 4.8$ for BW. Loss of stability at larger time steps is due to flux linearization and/or factorization errors, which prevent unconditional stability in practical computations~\citep{hirsch_07}. The time behavior of turbulence kinetic energy (panel (a)) is well predicted at all Courant numbers up to the stability limit, whereas pressure fluctuations (panel (b)) are overdamped starting at $\mathrm{CFL} \approx 3$, in both ATI and BW. The different behavior is caused by the fact that pressure receives contributions of both hydrodynamic and acoustic nature. As seen in the previous Section, acoustic waves undergo significant damping at high Courant number. This is even clearer in the velocity and pressure spectra, shown in panels (c) and (d), respectively. While velocity spectra are perfectly captured at all Courant numbers, pressure spectra undergo numerical damping, especially at intermediate wavenumbers, which is easily understood based on the amplification factors shown in Fig.~\ref{fig:gfac}. Given the similar performance of the two implicit methods for this test case, ATI is certainly preferable owing to its lower computational cost, which allows to achieve an effective speed-up over the explicit case (see table \ref{tab:cost}) of about a factor of three, whereas BW yields almost the same efficiency.

\subsection{Turbulent flow in plane channel} \label{sec:channel}

Channel flow is the simplest prototype of
wall-bounded flows, and it has been studied by many authors 
in the incompressible~\citep{kim_87, bernardini_14, lee_15}, as well as in
the compressible regime~\citep{coleman_95, lechner_01, modesti_16}.
The controlling parameters are the bulk Mach number
$M_b=u_b/c_w=1.5$ (where $u_b$ is the average velocity across the channel thickness, and 
$c_w$ the sound speed at the wall temperature), and the bulk Reynolds number $Re_b=2\rho_b u_bh/\mu_w=6000$ (where $\rho_b$ is the bulk density, $\mu_w$ the dynamic viscosity at the wall, and $h$ the channel half height).
All DNS are initialized with a parabolic velocity profile with
superposed small perturbations, whereas density and pressure are uniform.
Periodic boundary conditions are applied in the streamwise ($x$) and spanwise ($z$) 
coordinate directions, and no-slip, isothermal boundary conditions are applied at the walls.
A spatially uniform forcing is applied to the streamwise momentum equation, and dynamically 
adjusted in time to maintain constant mass flow rate~\citep{modesti_16}.
Favre density-weighted decomposition is applied to separate mean values from fluctuations, namely
$\phi=\widetilde{\phi} + \phi''$, with $\widetilde{\phi}=\overline{\rho\phi}/\overline{\rho}$).
\begin{table}
 \centering
 \begin{tabular}{lccccccccccccc}
 \hline
 Case & $M_b$ & $M_0$ & $Re_b$ & $Re_{\tau}$ & $\Delta y_w^+$ & $\Delta x^+$ & $\Delta z^+$ & $\Delta t_x^+$ & $\Delta t_y^+$ & $\Delta t_z^+$ & $\Delta t_{yv}^+$ & $\Delta t^+$ & CPU \\
 \hline
 CH01-EXPL      & 0.1 & 0.1  & 5790 & 180 & 0.60 & 8.80 & 3.90 & 0.053 &  0.0077  & 0.026 & 1.3  &  0.0077 & 1 \\
 CH01-ATI-XYZ   & 0.1 & 0.1  & 5790 & 180 & 0.60 & 8.80 & 3.90 & 0.053 &  0.0077  & 0.026 & 1.3  &  0.077  & 0.15 \\
 CH01-BW-XYZ    & 0.1 & 0.1  & 5790 & 180 & 0.60 & 8.80 & 3.90 & 0.053 &  0.0077  & 0.026 & 1.3  &  0.077  & 0.82 \\
 \hline
 CH15a-EXPL     & 1.5 & 1.28 & 6000 & 220 & 0.70 & 10.8 & 4.80 & 0.32  &  0.11   & 0.27  & 1.2  &  0.099  & 1 \\
 CH15a-ATI-Y    & 1.5 & 1.28 & 6000 & 220 & 0.70 & 10.8 & 4.80 & 0.32  &  0.11   & 0.27  & 1.2  &  0.24  & 0.48 \\
 CH15a-BW-Y     & 1.5 & 1.28 & 6000 & 220 & 0.70 & 10.8 & 4.80 & 0.32  &  0.11   & 0.27  & 1.2  &  0.24  & 0.70 \\
 \hline
 CH15b-EXPL     & 1.5 & 1.28 & 6000 & 220 & 0.15 & 10.8 & 4.08 & 0.32  &  0.11  & 0.27  & 0.062  &  0.021 & 1 \\
 CH15b-AVTI-Y   & 1.5 & 1.28 & 6000 & 220 & 0.15 & 10.8 & 4.80 & 0.32  &  0.11  & 0.27  & 0.062  &  0.21 & 0.13  \\
 CH15b-BWV-Y    & 1.5 & 1.28 & 6000 & 220 & 0.15 & 10.8 & 4.80 & 0.32  &  0.11  & 0.27  & 0.062  &  0.21 & 0.19  \\
 \hline
 \end{tabular}
 \caption{Flow parameters for DNS of plane channel flow (CH). 
   $M_b$ and $Re_b$ are the bulk Reynolds and Mach number, respectively.
$M_0 = M_b \sqrt{T_w/T_b}$ is the reference Mach number, introduced when discussing Eqn.~\eqref{eq:dtinv}.
The computational box dimension is $4\pi h\times2h\times4/3\pi$ for all flow cases.
$\Delta y_w^+$ is the distance of the first grid point from the wall, and
$\Delta x^+$, $\Delta z^+$ are the streamwise and spanwise grid spacings.
The $\Delta t_i^+$ are the allowable time steps in the coordinate directions, estimated according to Eqns.~\eqref{eq:dtinv},\eqref{eq:dtvis}.
 $\Delta t^+$ is the time step actually used in the simulations.
CPU is the cost to cover a unit time interval, compared to the standard fully explicit algorithm (EXPL).}
\label{tab:channel}
\end{table}

The main flow parameters are listed in Tab.~\ref{tab:channel}.
Three flow cases have been considered,
one at $M_b=0.1$ (denoted as CH01), and 
two at $M_b=1.5$ (denoted as CH15a-b), 
the latter two only differing in the distance of the first grid point from the wall.
Reference DNS have been carried out with fully explicit time discretization, at $\mathrm{CFL} \approx 1$,
which are used as a basis of comparison for the ATI and BW algorithms.
In order to understand the effectiveness of the (semi-)implicit algorithms, in Tab.~\ref{tab:channel} we report the
time step restrictions associated with the three coordinate directions, as estimated from Eqns.~\eqref{eq:dtinv},\eqref{eq:dtvis},
as well as the actual time step used in the DNS, all in wall units.
As expected, in all flow cases the time step limitation in the wall-normal direction is the most restrictive.
Although larger time steps are allowed on grounds of sole numerical stability, 
all DNS have been carried out at the maximum time step for which accurate results are obtained,
which corresponds to $CLF \approx 1$ for the fully explicit simulations.
For ease of reference, the maximum time steps associated with accuracy and stability restrictions are
also reported in Fig.~\ref{fig:dtconv}(a) with circle and square symbols, respectively.

As a first test, we consider flow at low subsonic Mach number (CH01), for which the explicit time advancement step
is very small, hence we apply implicit treatment is all coordinate directions (XYZ). 
We find that, although the wall-normal time step restrictions can be removed, the allowed time step for accurate
calculations cannot substantially larger than for the streamwise convective restriction (see Fig.~\ref{fig:dtconv}(a)). This is probably due to
inherent mesh anisotropy in DNS of wall-bounded flows. In fact, mesh spacing is over-resolved in the wall-normal direction, hence
the relevant values of the reduced wavenumber $k h$ are small, which allows to operate at high values
of $\mathrm{CFL}$ with little error, recalling (see Fig.~\ref{fig:gfac}) that the dissipation error grows with both $k h$ and $\mathrm{CFL}$. 
On the other hand, the typical wall-parallel mesh spacings used in DNS are barely sufficient to resolve the smallest scales
of turbulence, hence the typical reduced wavenumbers are higher, and time accuracy is a factor in that case.
We find that both ATI and BW are capable of boosting the time step by about a factor of ten,
with efficiency gain of $85\%$ for ATI, and results almost indistinguishable from the fully explicit case (see below).
Still, the time step is far from that allowed by incompressible solvers (again, see Fig.~\ref{fig:dtconv}(a)).
This issue will be further recalled in the concluding discussion.

To show effectiveness in removing the wall-normal acoustic time limitation is supersonic flow calculations, 
in flow case CH15a the first grid point is placed sufficiently far from the wall that the viscous limitation is ineffective.
Hence, the implicit algorithms are applied only in the wall-normal direction (Y), and viscous terms are handled explicitly.
The ATI and BW algorithms are both found to effectively suppress the wall-normal acoustic time step limitation, 
and achieve the same maximum time step for accurate flow resolution,
corresponding to about $\mathrm{CFL} = 2.4$. Hence, accounting for the cost figures given in table~\ref{tab:cost},
we find a speed-up of about a factor of two for the ATI algorithm, and $30\%$ gain with BW.

To prove effectiveness of the implicit treatment of the viscous terms proposed in Section~\eqref{sec:viscous},
in flow case CH15b the first grid point is placed closer to the wall, in such a way that the viscous
time limitation also becomes relevant, after the acoustic one. Both wall-normal time step restrictions are suppressed
through use of the AVTI and BWV algorithms, hence the achieved time step is similar to flow case CH15a.
Both algorithms here achieve $\mathrm{CFL} \approx 10$, at a cost which is a small fraction of the fully explicit algorithm.

\begin{figure}
\centering
\psfrag{X}[t][][1.0]{$y^+$}
\psfrag{Y}[b][][1.0]{$u^+$}
(a)
\includegraphics[width=5cm,clip,angle=270]{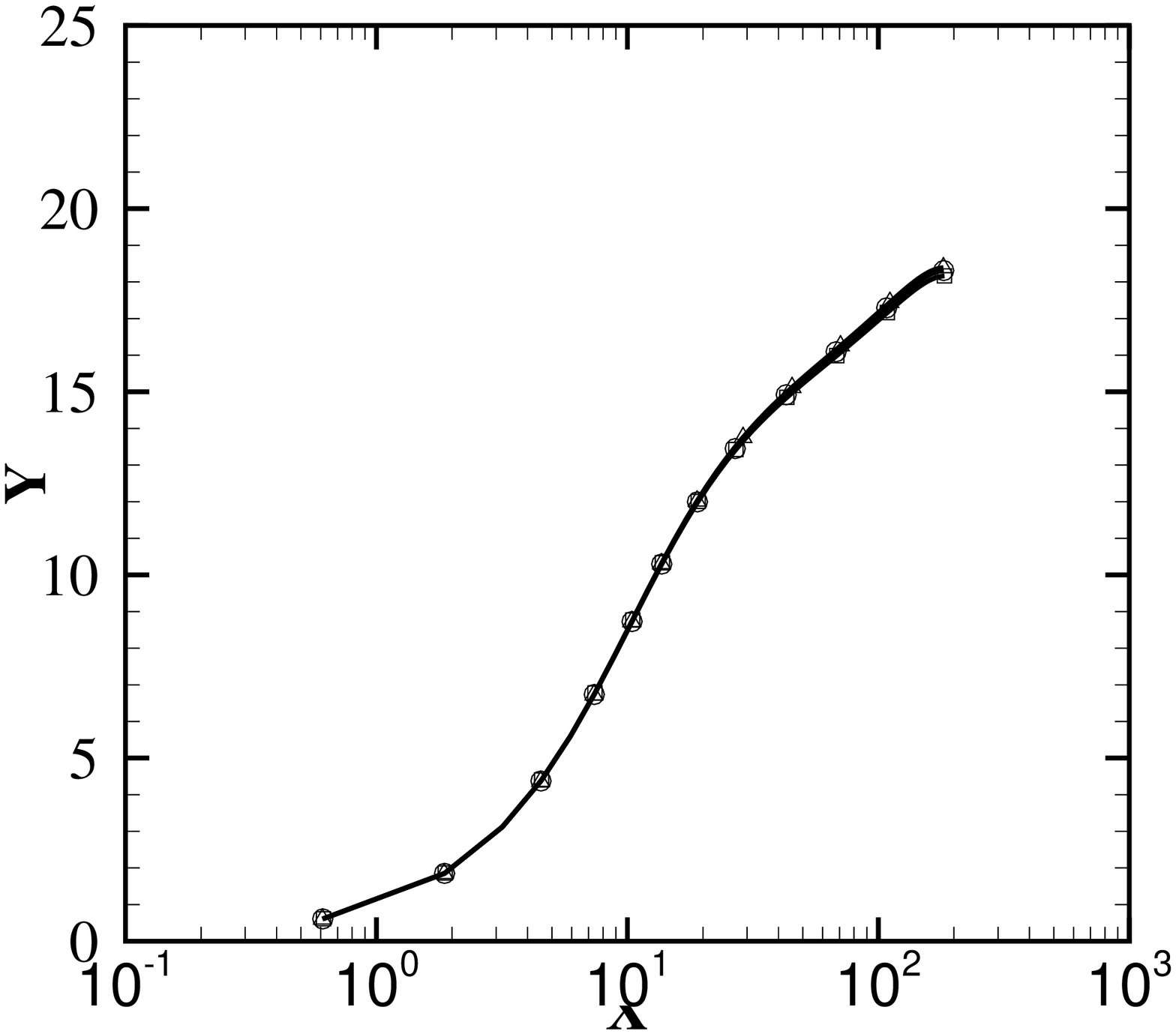}
\psfrag{X}[t][][1.0]{$y^+$}
\psfrag{Y}[b][][1.0]{$\tau_{ii}/\tau_w$}
(b)
\psfrag{a}[][][1.0]{$\tau_{11}$}
\psfrag{b}[][][1.0]{$\tau_{22}$}
\psfrag{c}[][][1.0]{$\tau_{33}$}
\includegraphics[width=5cm,clip,angle=270]{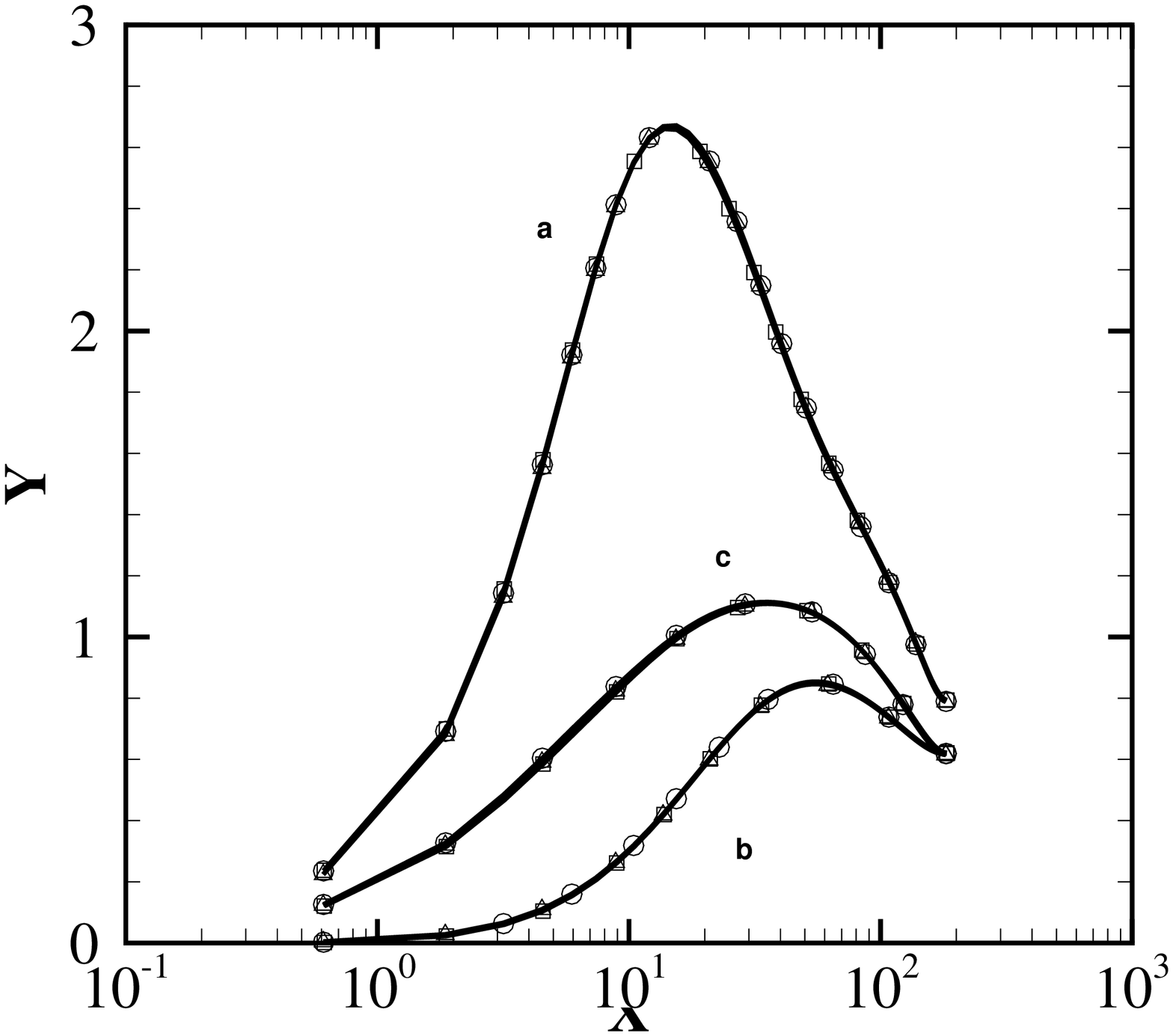}\\
\vskip1.em
\psfrag{X}[t][][1.0]{$y^+$}
\psfrag{Y}[b][][1.0]{$p_{rms}/{\tau_w}$}
(c)
\includegraphics[width=5cm,clip,angle=270]{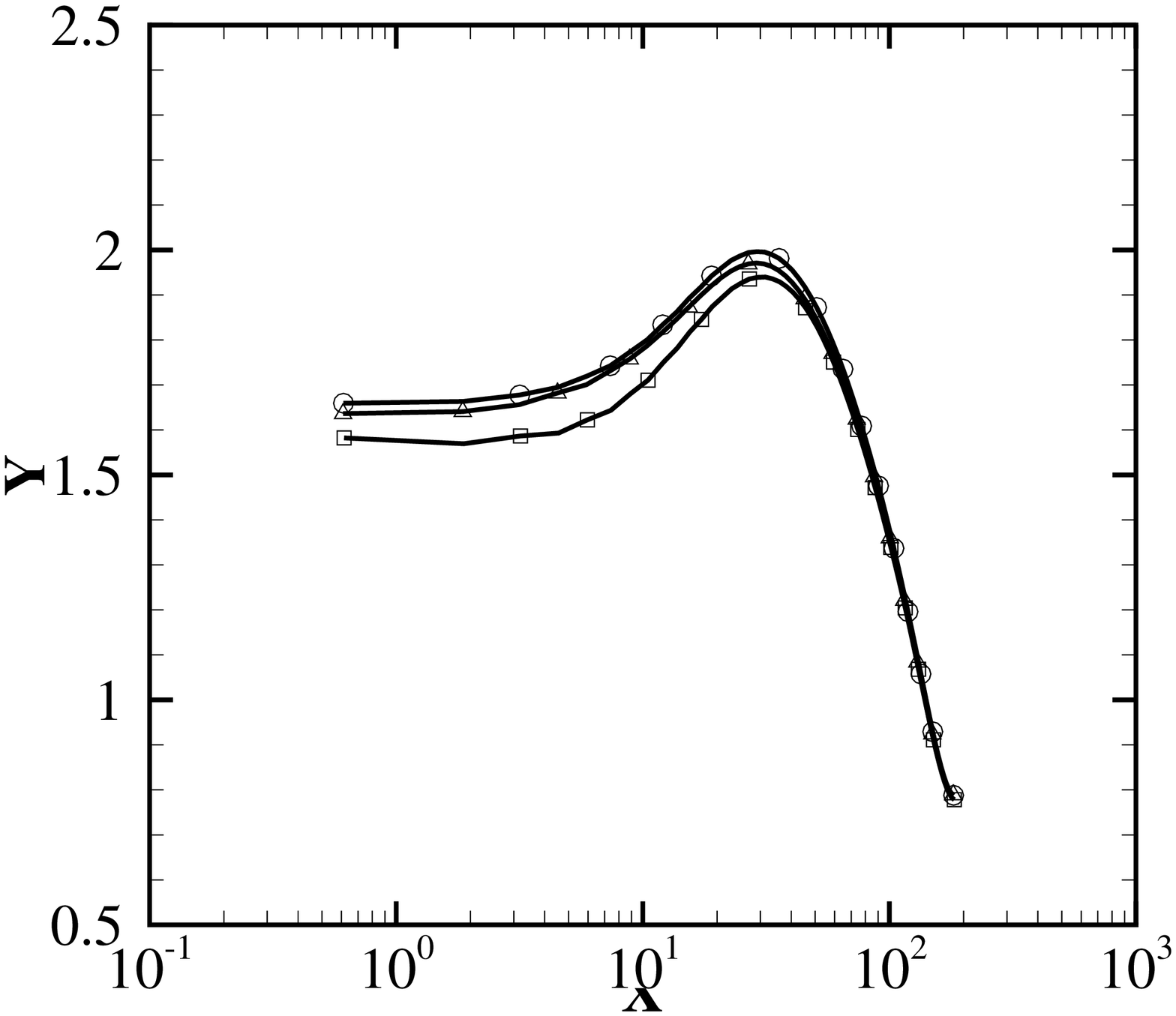}
\psfrag{X}[t][][1.0]{$y^+$}
\psfrag{Y}[b][][1.0]{$T_{rms}/T_{\tau}$}
(d)
\includegraphics[width=5cm,clip,angle=270]{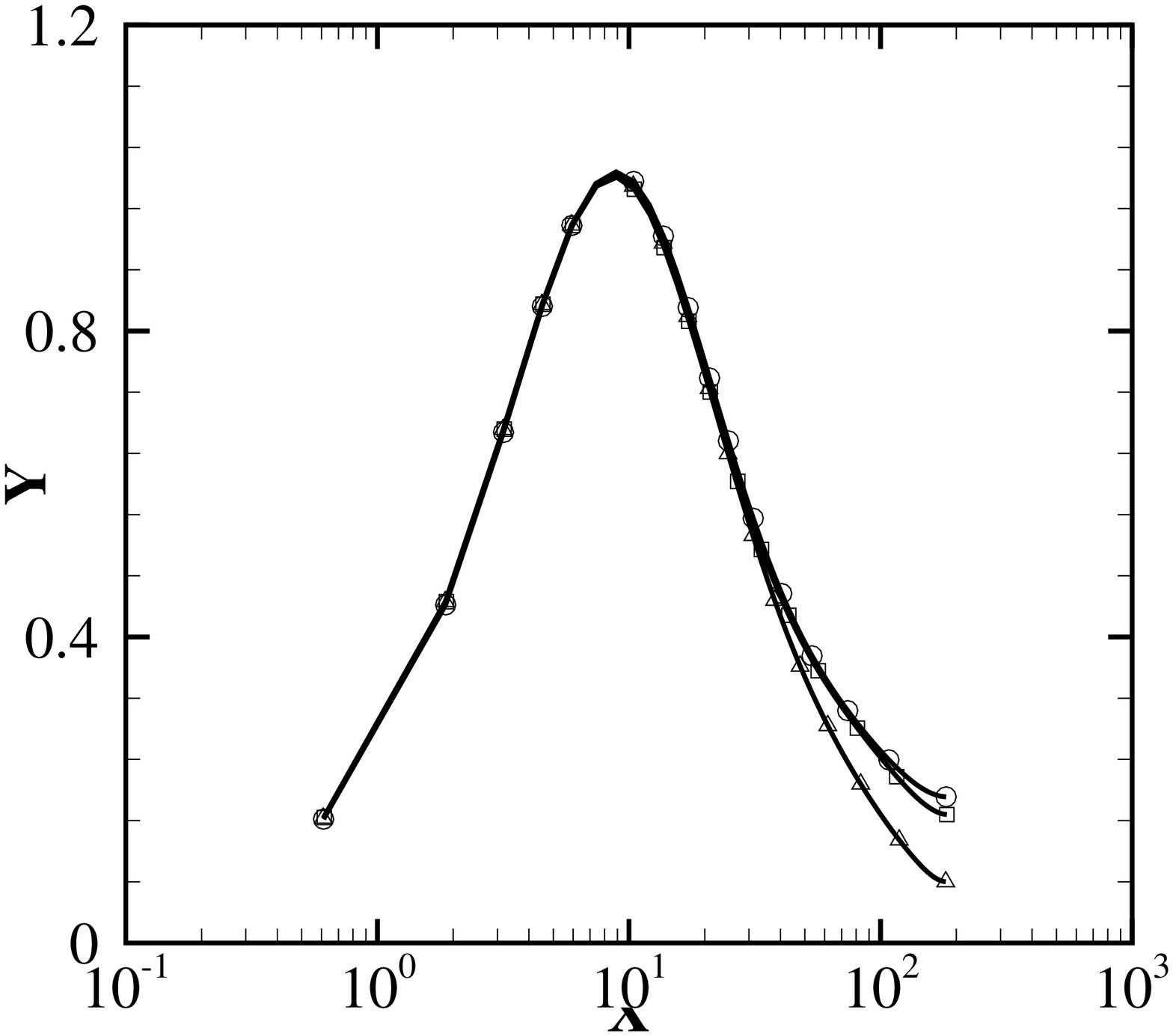}\\
\vskip1.em
\caption{Flow statistics for DNS of flow case CH01 (see Table~\ref{tab:channel}):
mean velocity (a), Reynolds stresses (b), r.m.s. pressure (c) and r.m.s. temperature (d),  
for CH01-EXPL (squares), CH01-ATI-XYZ (circles), CH01-BW-XYZ (triangles). $T_{\tau} = q_w / (\rho_w c_p u_{\tau})$ is the friction temperature.}
\label{fig:CH01}
\end{figure}

\begin{figure}
\centering
\psfrag{X}[t][][1.0]{$y^+$}
\psfrag{Y}[b][][1.0]{$u^+$}
(a)
\includegraphics[width=5cm,clip,angle=270]{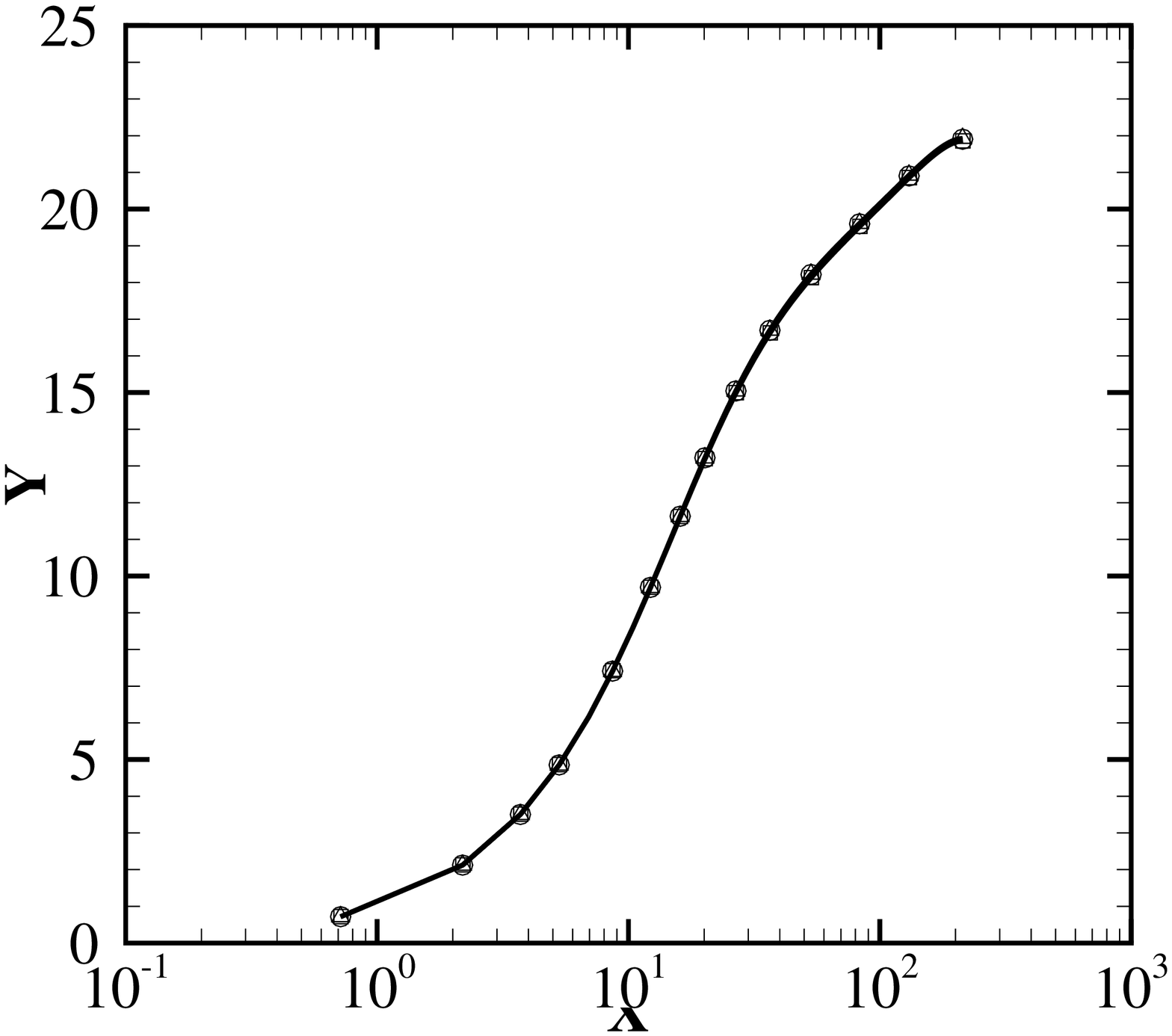}
\psfrag{X}[t][][1.0]{$y^+$}
\psfrag{Y}[b][][1.0]{$\tau_{ii}/\tau_w$}
(b)
\psfrag{a}[][][1.0]{$\tau_{11}$}
\psfrag{b}[][][1.0]{$\tau_{22}$}
\psfrag{c}[][][1.0]{$\tau_{33}$}
\includegraphics[width=5cm,clip,angle=270]{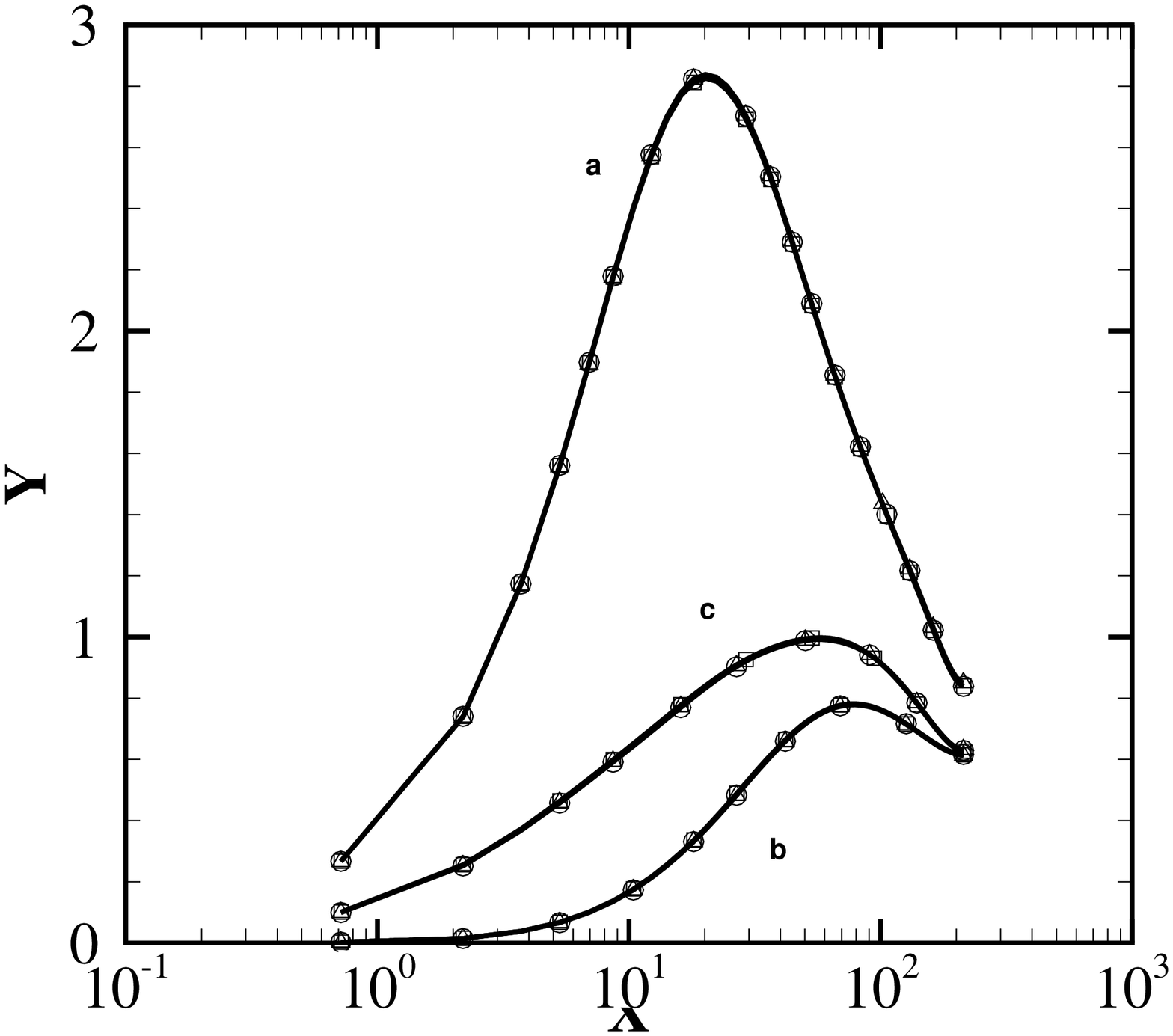}\\
\vskip1.em
\psfrag{X}[t][][1.0]{$y^+$}
\psfrag{Y}[b][][1.0]{$p_{rms}/{\tau_w}$}
(c)
\includegraphics[width=5cm,clip,angle=270]{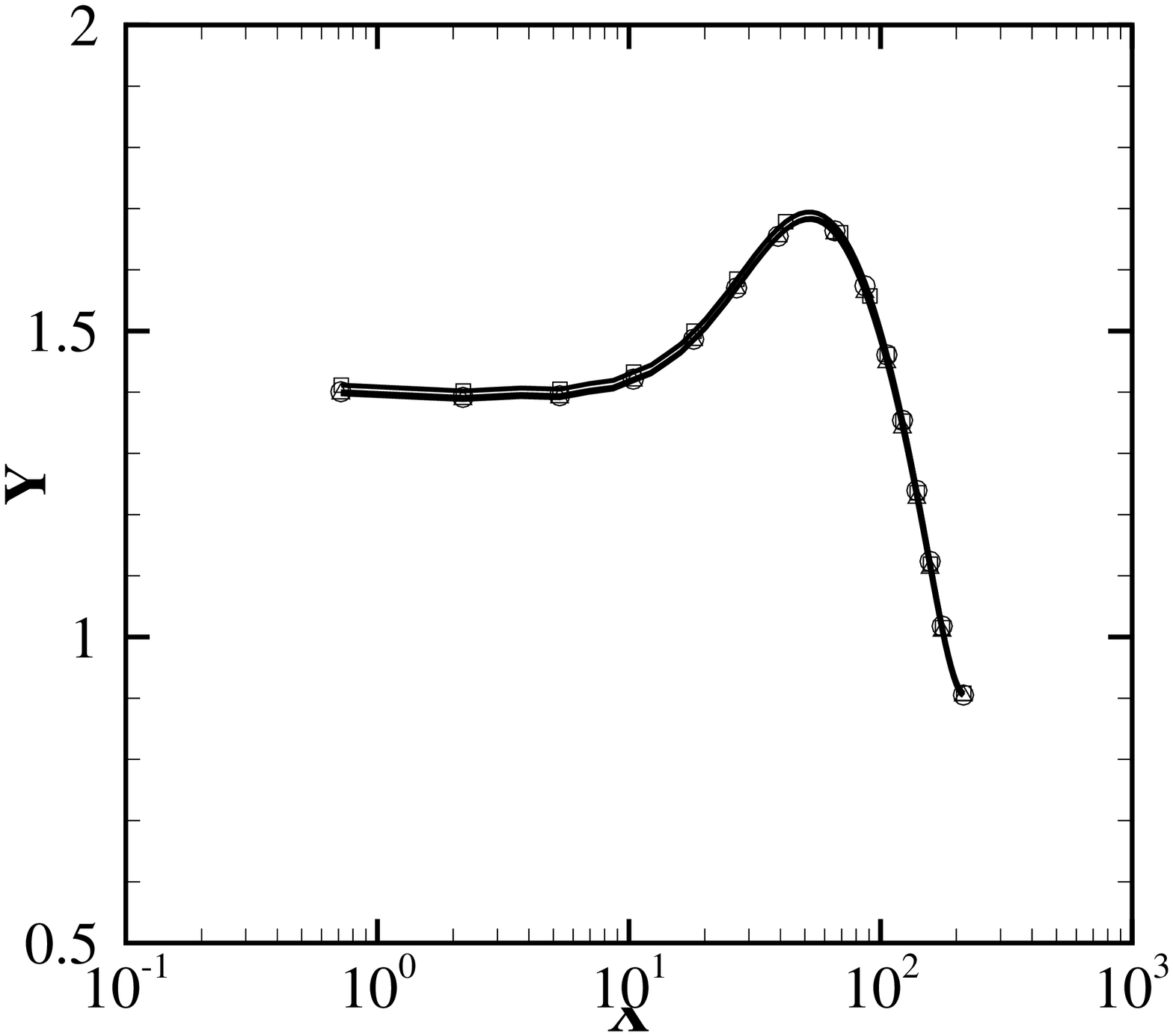}
\psfrag{X}[t][][1.0]{$y^+$}
\psfrag{Y}[b][][1.0]{$T_{rms}/T_{\tau}$}
(d)
\includegraphics[width=5cm,clip,angle=270]{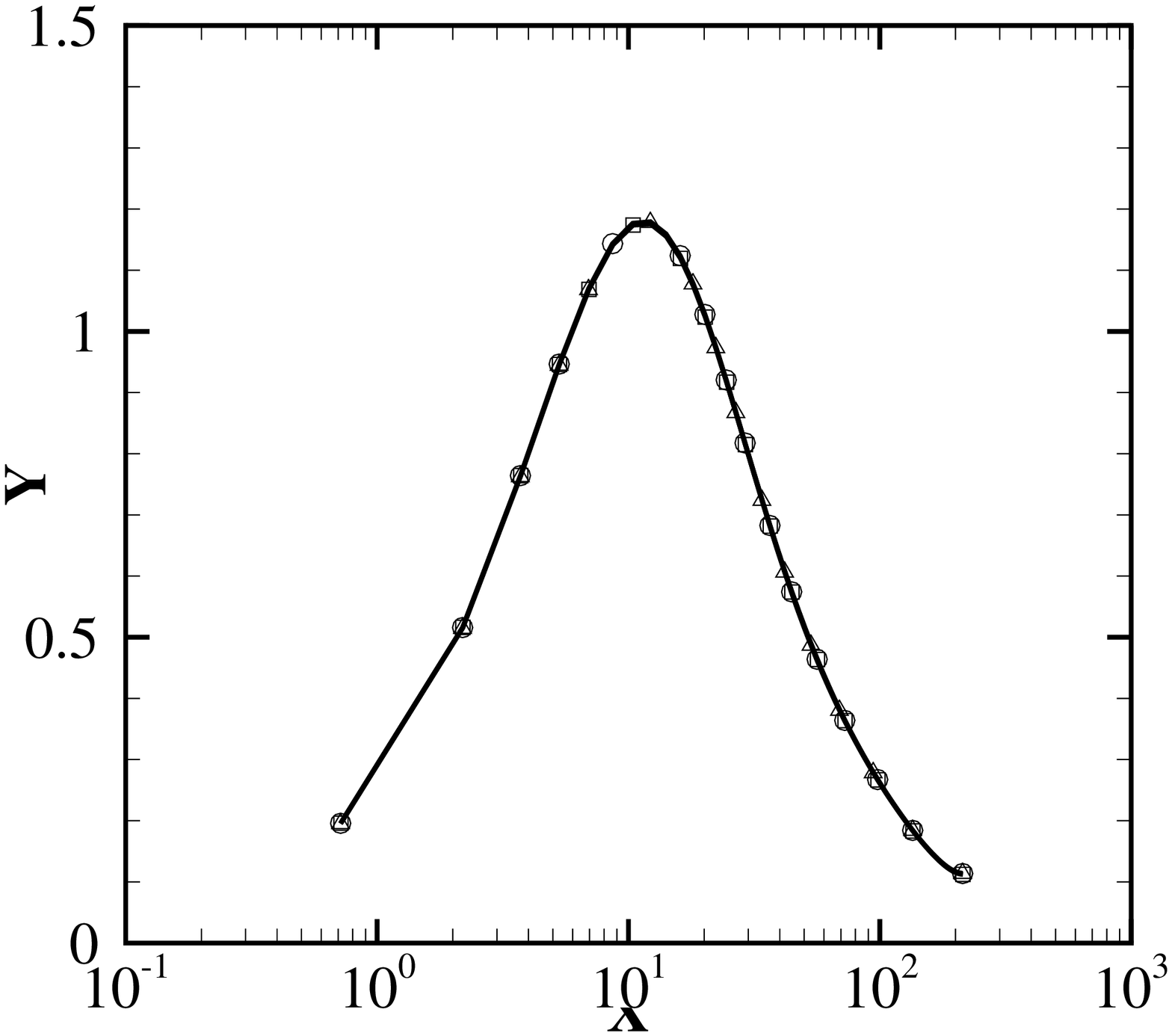}\\
\vskip1.em
\caption{Flow statistics for DNS of flow case CH15a (see Table~\ref{tab:channel}):
mean velocity (a), Reynolds stresses (b), r.m.s. pressure (c) and r.m.s. temperature (d),  
for CH15a-EXPL (squares), CH15a-ATI-Y (circles), CH15a-BW-Y (triangles).}
\label{fig:CH15a}
\end{figure}

\begin{figure}
\centering
\psfrag{X}[t][][1.0]{$y^+$}
\psfrag{Y}[b][][1.0]{$u^+$}
(a)
\includegraphics[width=5cm,clip,angle=270]{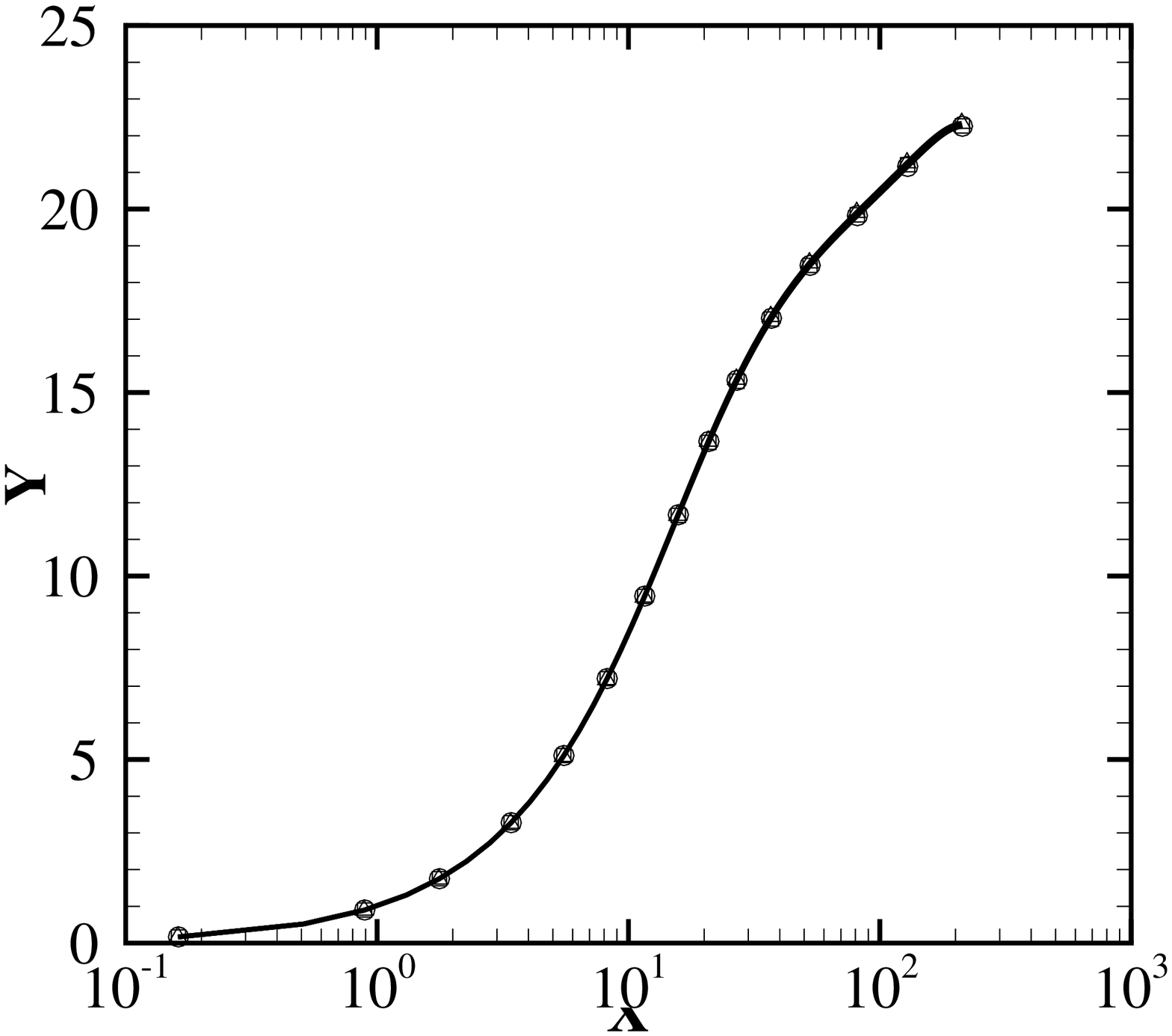}
\psfrag{X}[t][][1.0]{$y^+$}
\psfrag{Y}[b][][1.0]{$\tau_{ii}/\tau_w$}
(b)
\psfrag{a}[][][1.0]{$\tau_{11}$}
\psfrag{b}[][][1.0]{$\tau_{22}$}
\psfrag{c}[][][1.0]{$\tau_{33}$}
\includegraphics[width=5cm,clip,angle=270]{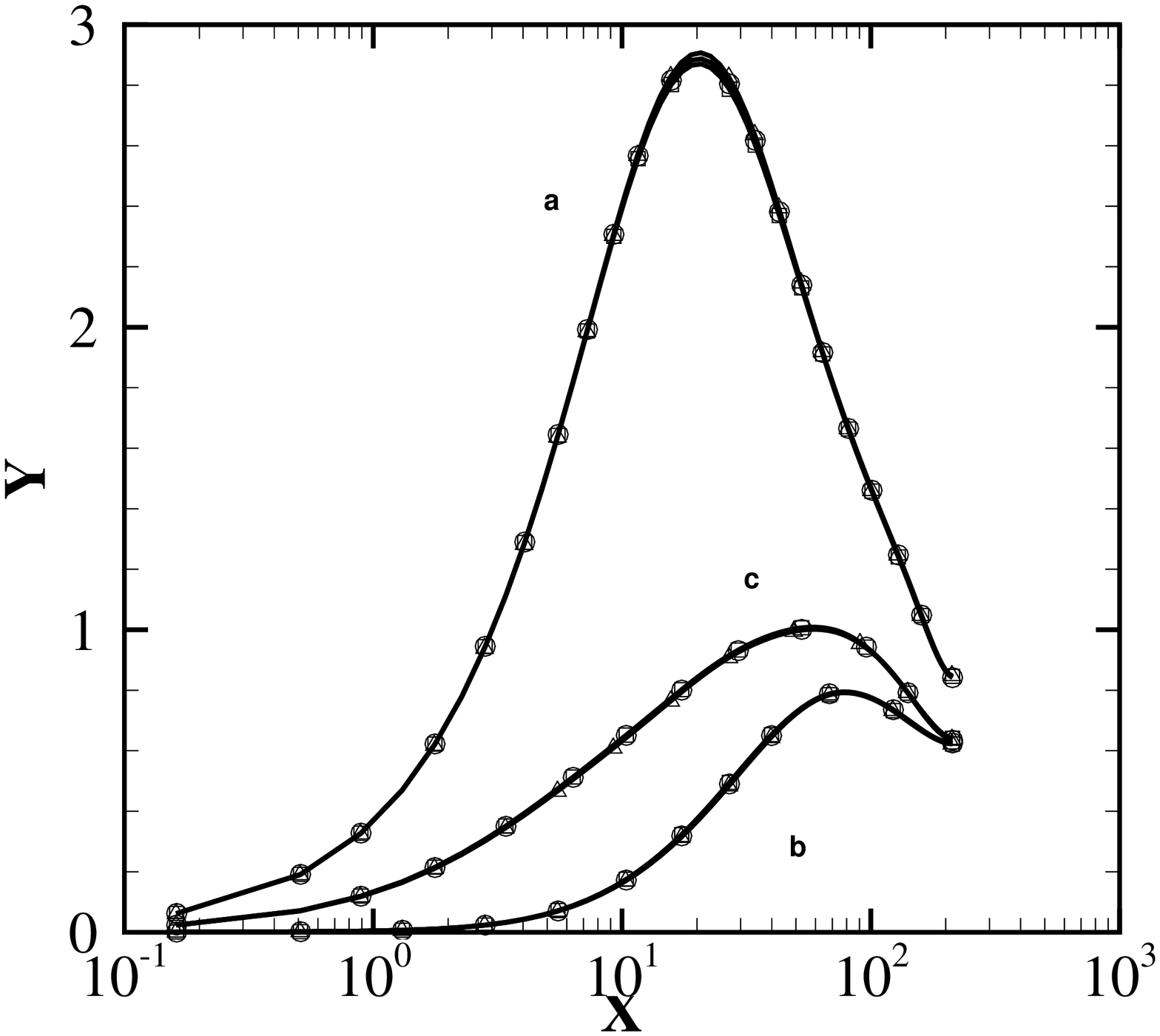}\\
\vskip1.em
\psfrag{X}[t][][1.0]{$y^+$}
\psfrag{Y}[b][][1.0]{$p_{rms}/{\tau_w}$}
(c)
\includegraphics[width=5cm,clip,angle=270]{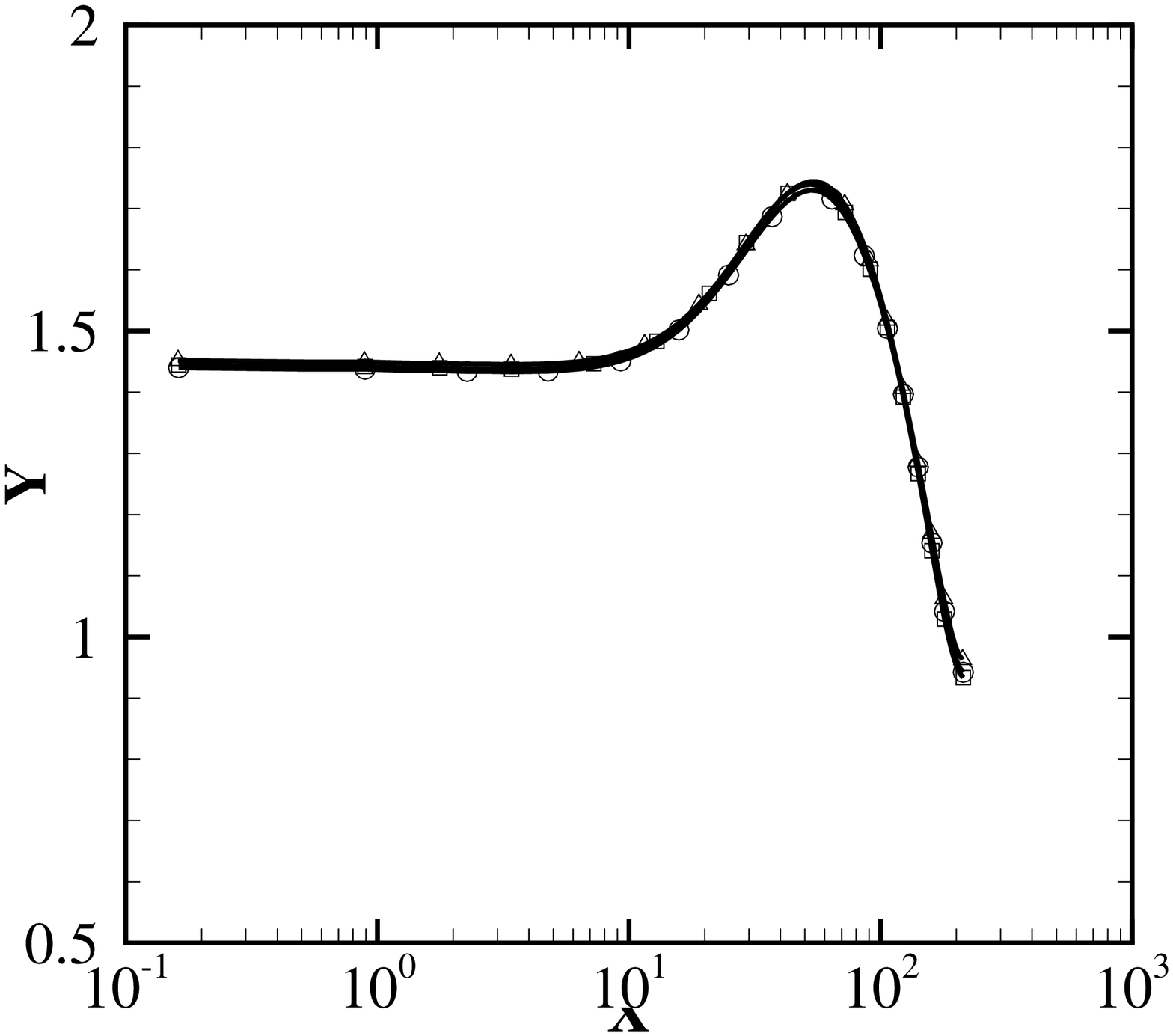}
\psfrag{X}[t][][1.0]{$y^+$}
\psfrag{Y}[b][][1.0]{$T_{rms}/T_{\tau}$}
(d)
\includegraphics[width=5cm,clip,angle=270]{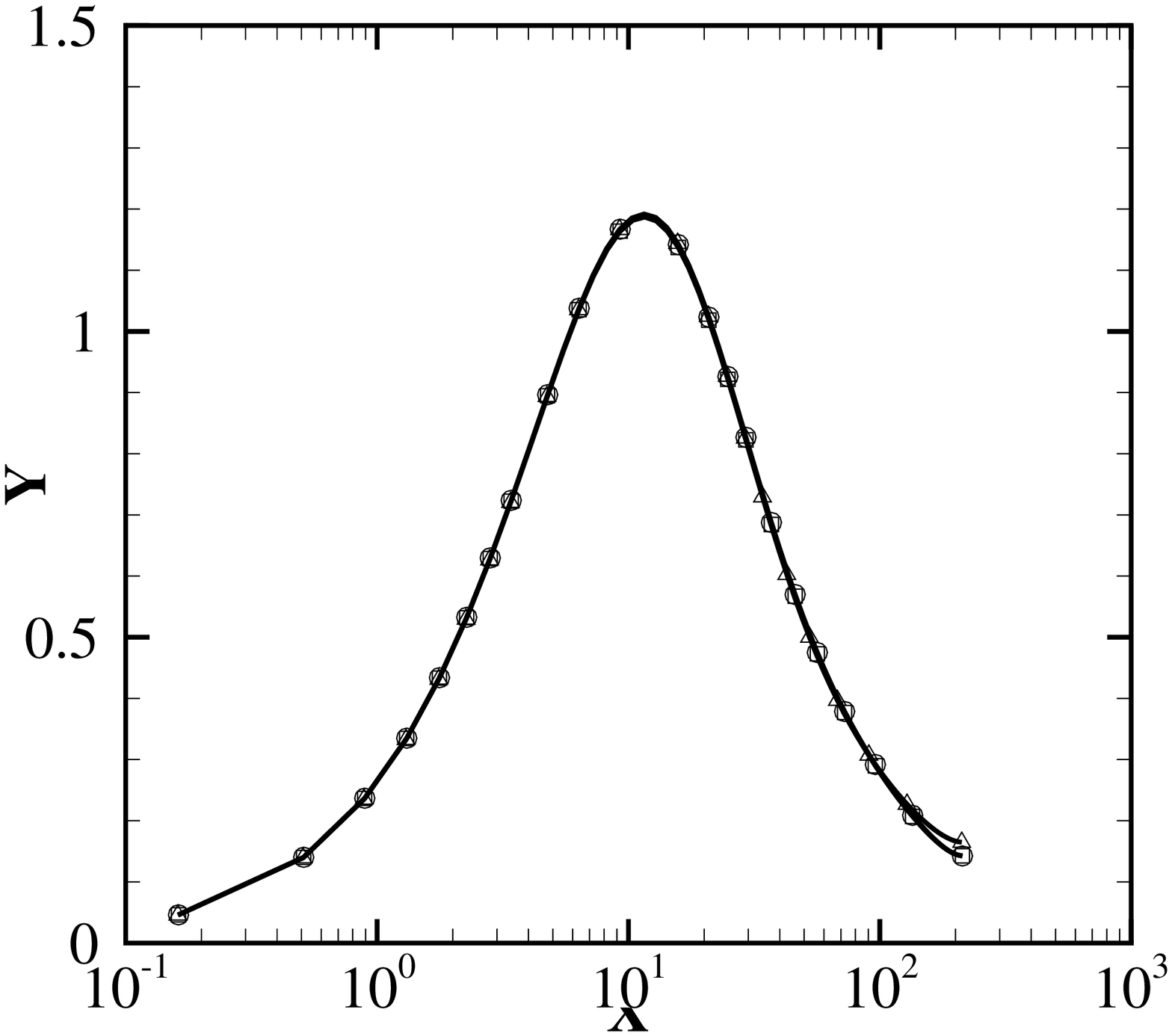}\\
\vskip1.em
\caption{Flow statistics for DNS of flow case CH15b (see Table~\ref{tab:channel}):
mean velocity (a), Reynolds stresses (b), r.m.s. pressure (c) and r.m.s. temperature (d),  
for CH15b-EXPL (squares), CH15b-AVTI-Y (circles), CH15b-BWV-Y (triangles).}
\label{fig:CH15b}
\end{figure}

For the sake of comparison, in Figs.~\ref{fig:CH01}-\ref{fig:CH15b} we show the main statistics 
for the flow cases listed in Table~\ref{tab:channel}.
As anticipated, excellent agreement is observed between implicit algorithms and the reference explicit solution, including
pressure and temperature fluctuations, which is especially satisfactory.
\subsection{Turbulent flow in square duct} \label{sec:duct}
\begin{table}
\centering
\begin{tabular}{lccccccccccccc}
\hline
Case & $M_b$ & $M_0$ & $Re_b$ & $Re_{\tau}$ & $\Delta y_w^+$ & $\Delta x^+$ & $\Delta z^+$ & $\Delta t_x^+$ & $\Delta t_y^+$ & $\Delta t_z^+$ & $\Delta t_{yv}^+$ & $\Delta t^+$ & CPU \\
\hline
DU02-EXPL    & 0.2 & 0.2 & 4410 & 150 & 0.66 & 8.40 &  0.66-3.20   & 0.094  &  0.019  &  0.019  &  1.69  &   0.018  &  1 \\
DU02-ATI-XYZ & 0.2 & 0.2 & 4410 & 150 & 0.66 & 8.40 &  0.66-3.20   & 0.094  &  0.019  &  0.019  &  1.69  &   0.18   &  0.15 \\
\hline
\end{tabular}
\caption{DNS dataset for square duct (DU) flow. 
$M_b$ and $Re_b$ are the bulk Reynolds and Mach number, respectively.
$M_0 = M_b \sqrt{T_w/T_b}$ is the reference Mach number, introduced when discussing Eqn.~\eqref{eq:dtinv}.
The computational box dimension is $8\pi h \times 2h \times 2h$.
$\Delta y_w^+$ is the distance of the first grid point from the wall, and
$\Delta x^+$,$\Delta z^+$ are the streamwise and spanwise grid spacings.
The $\Delta t_i^+$ are the allowable time steps in the coordinate directions, estimated according to Eqns.~\eqref{eq:dtinv},\eqref{eq:dtvis}.
$\Delta t^+$ is the time step actually used in the simulations.
CPU is the cost to cover a unit time interval, compared to the standard fully explicit algorithm (EXPL).}
\label{tab:duct}
\end{table}

\begin{figure}
\centering
\psfrag{X}[t][][1.0]{$y^+$}
\psfrag{Y}[b][][1.0]{$u^+$}
(a)
\includegraphics[width=5cm,clip,angle=270]{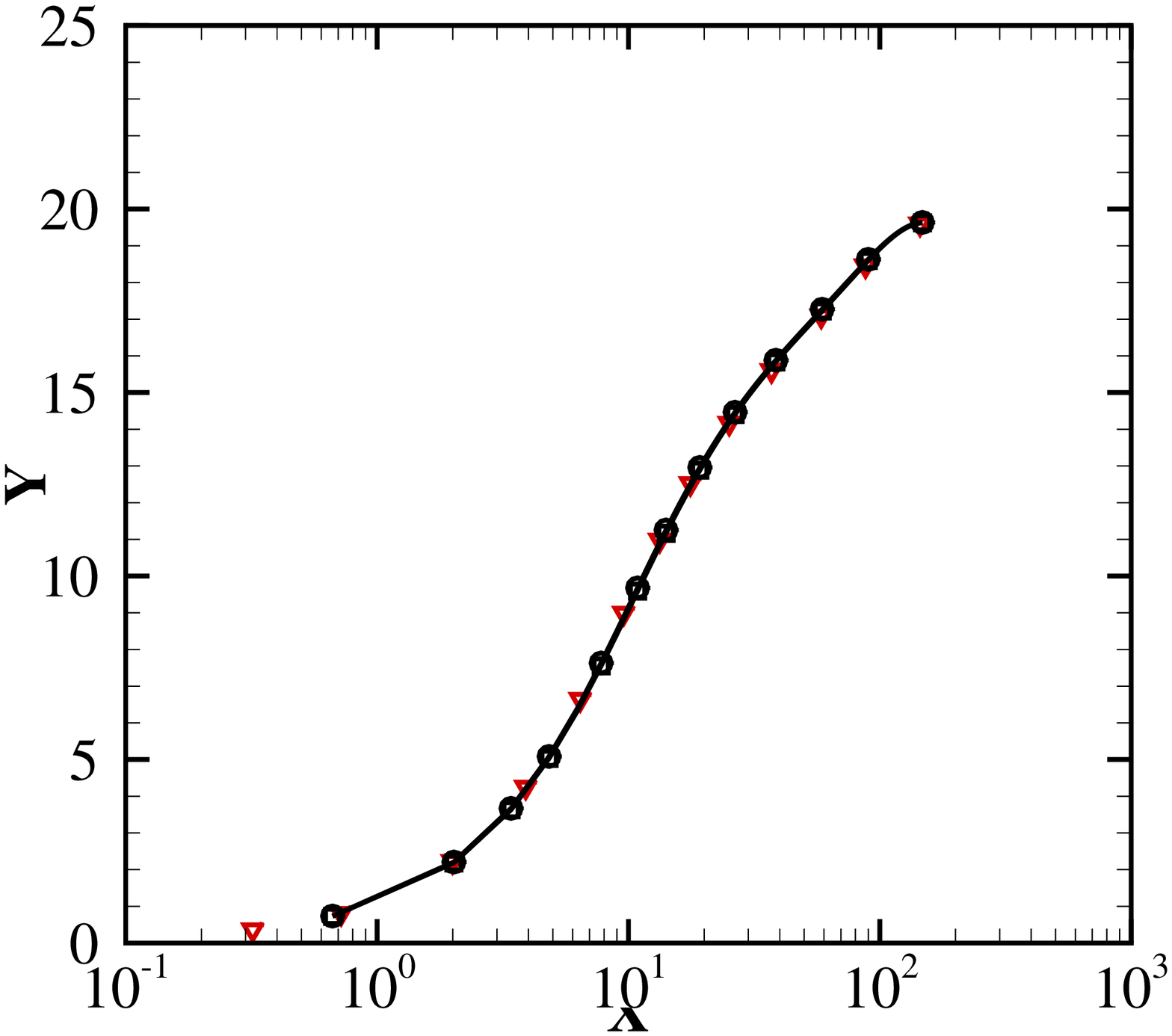}
\psfrag{X}[t][][1.0]{$y^+$}
\psfrag{Y}[b][][1.0]{$\tau_{ii}/\tau_w$}
(b)
\psfrag{a}[][][1.0]{$\tau_{11}$}
\psfrag{b}[][][1.0]{$\tau_{22}$}
\psfrag{c}[][][1.0]{$\tau_{33}$}
\includegraphics[width=5cm,clip,angle=270]{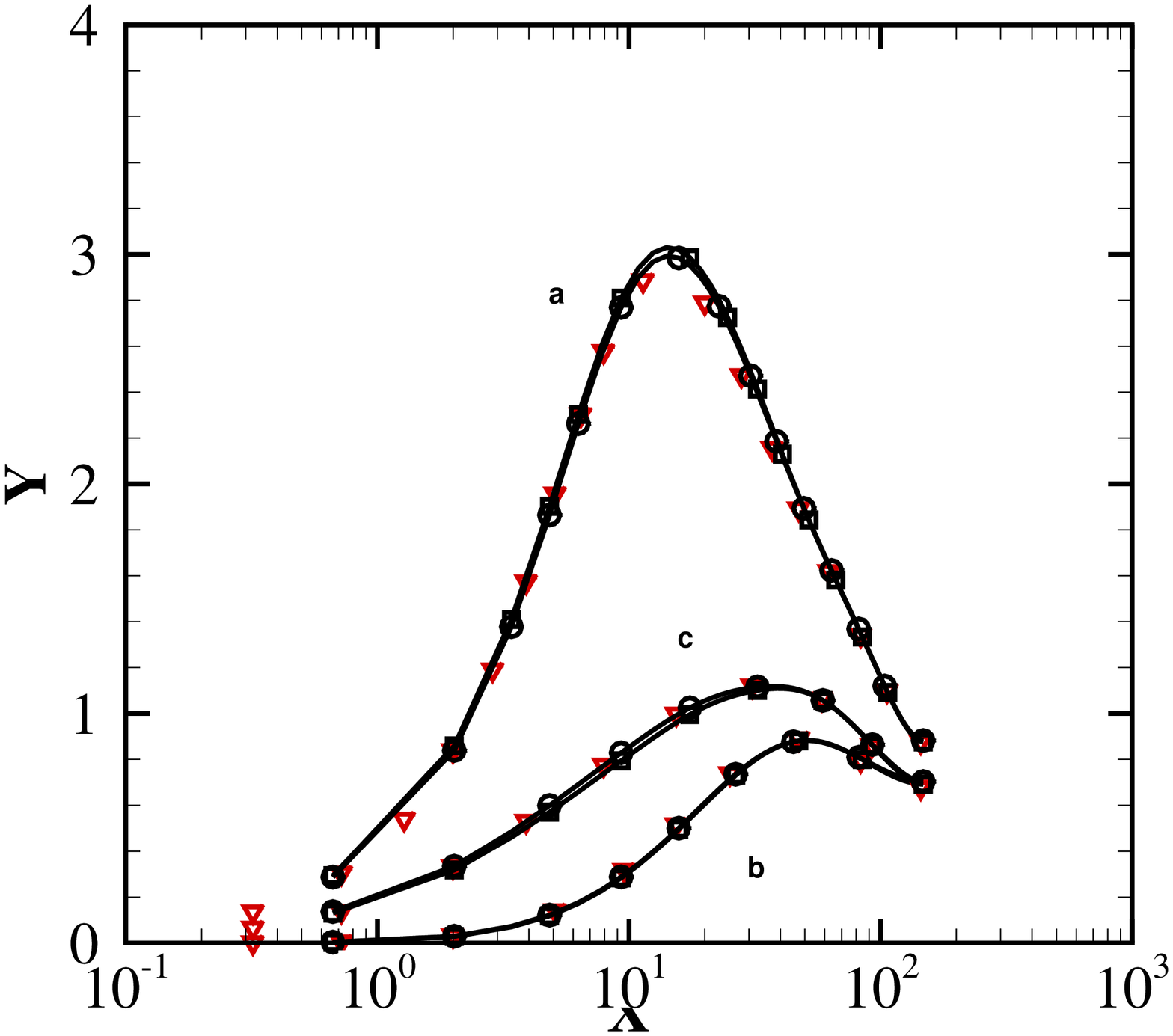}\\
\psfrag{X}[t][][1.0]{$y^+$}
\psfrag{Y}[b][][1.0]{$p_{rms}/{\tau_w}$}
(c)
\includegraphics[width=5cm,clip,angle=270]{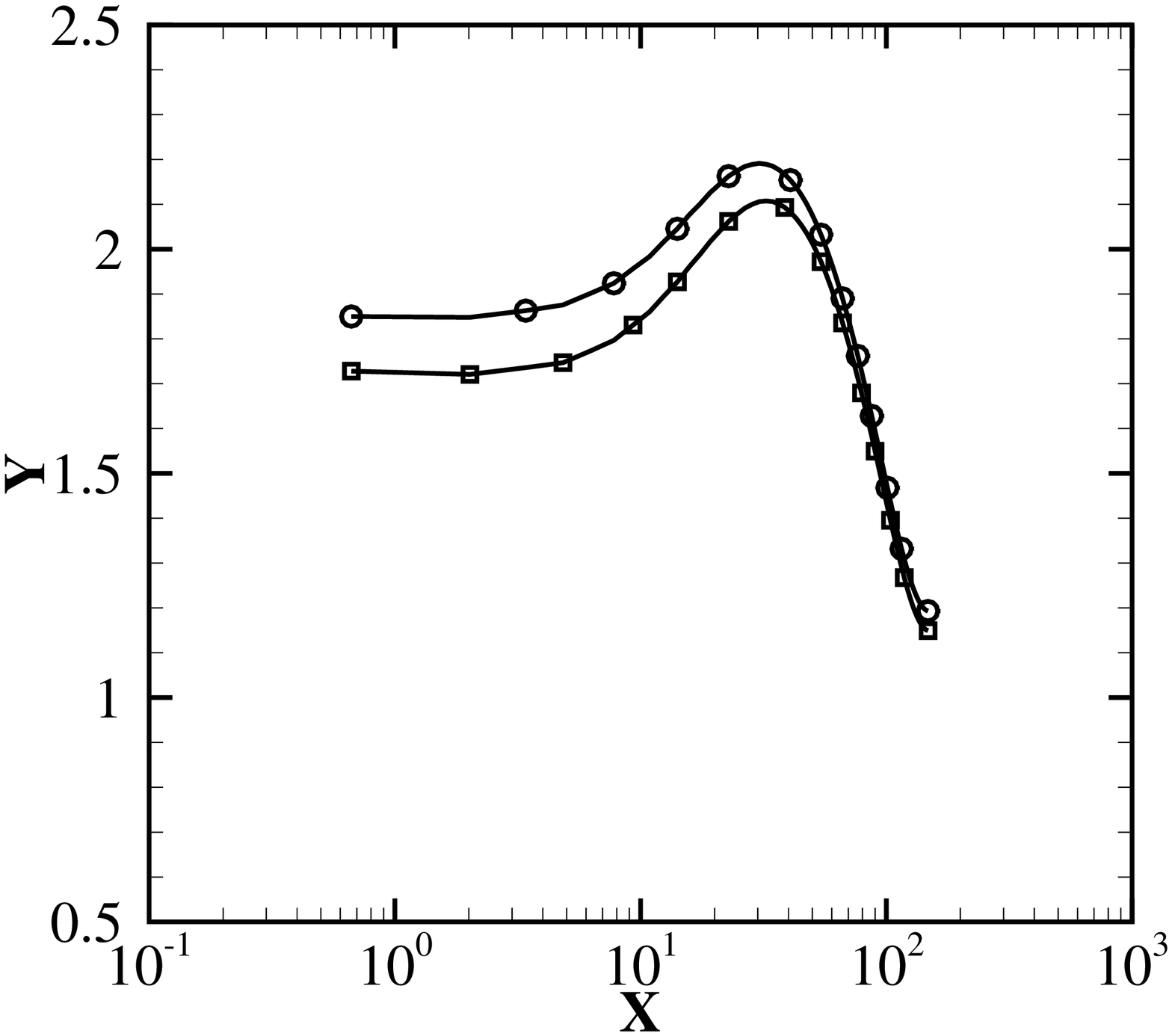}
\psfrag{X}[t][][1.0]{$y^+$}
\psfrag{Y}[b][][1.0]{$T_{rms}/T_{\tau}$}
(d)
\includegraphics[width=5cm,clip,angle=270]{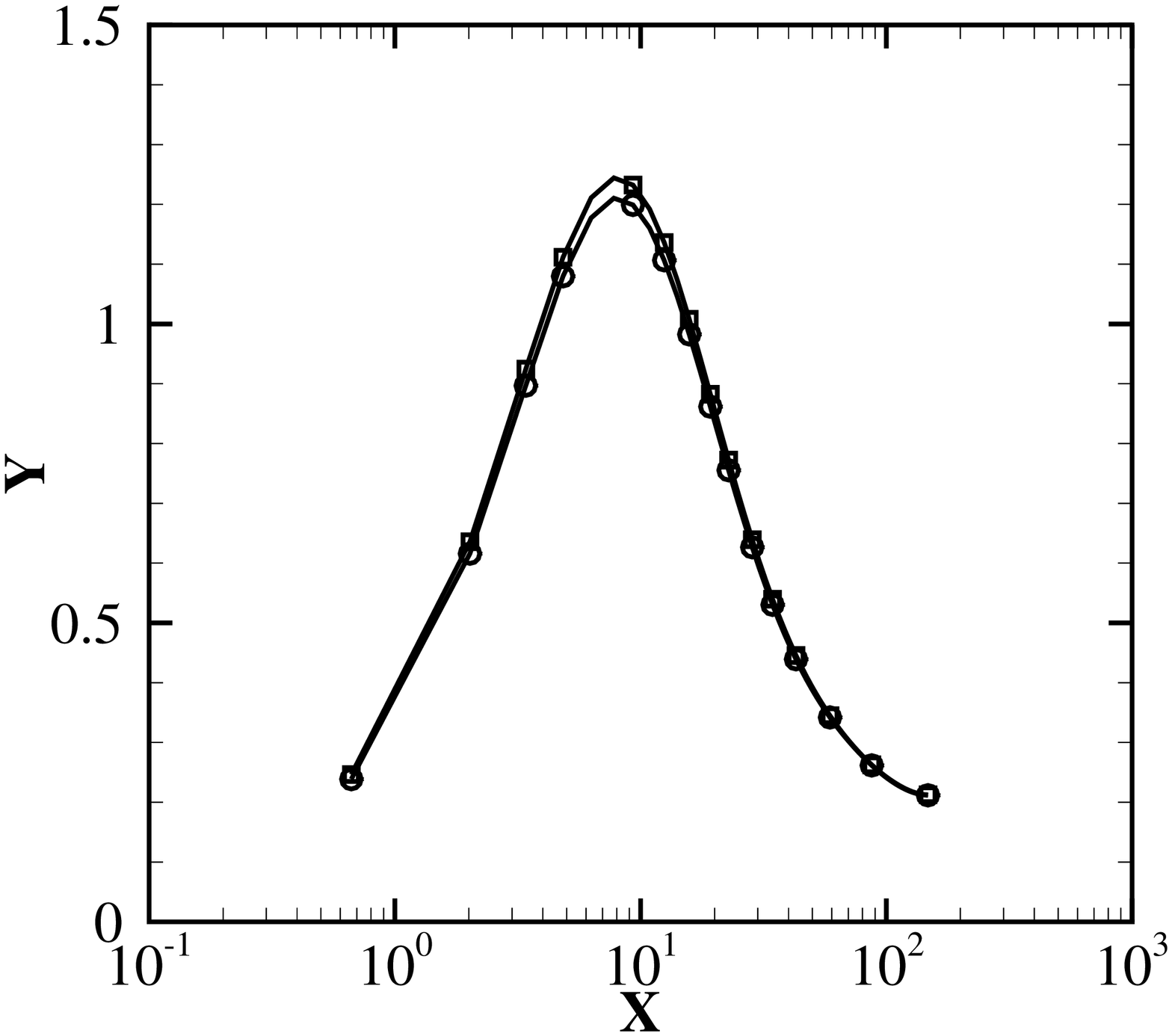}\\
\caption{DNS of flow in square duct (see Table~\ref{tab:duct}):
mean velocity (a), Reynolds stresses (b), r.m.s. pressure (c) and r.m.s. temperature (d),  
for DU02-EXPL (squares), DU02-ATI-XYZ (circles). Triangle symbols denote reference incompressible DNS data~\citep{pinelli_10}.}
\label{fig:duct}
\end{figure}

As a further step in complexity we consider the flow inside a straight
duct with square cross-section. This flow has been the subject of several DNS studies
in the incompressible regime~\citep{gavrilakis_92,huser_93,pinelli_10}, all limited
to low Reynolds number. 
One of the main difficulties that arise when dealing with square duct
flows is the long averaging time necessary to attain convergence of even the basic
mean flow statistics, caused by the extremely long typical time scales of secondary corner eddies. 
In fact, \citet{pinelli_10} reported that an averaging time of about $8000 h/u_b$ was needed
to have symmetric statistics in the four quadrants of the cross section.
Hence, it is clear that efficient numerical methods are needed to study turbulent compressible flow in ducts.
Numerical simulations have been here carried out (see Table~\ref{tab:duct} for the main flow parameters) 
at the same Reynolds number as \citet{pinelli_10},
and sufficiently low Mach number ($M_b=0.2$) that direct comparison with incompressible data is possible.
The duct length $L_x=8h$ (where $2h$ is the length of each side of the duct), and the time window for 
collecting the flow statistics is the same used by \citet{pinelli_10}.
As in plane channel flow, a spatially uniform forcing is applied to the momentum equation to maintain
a time constant mass flow rate. Note that, unlike in channel flow, the mesh is also non-uniformly spaced in
the $z$ direction, hence a range of mesh spacings is reported in Table~\ref{tab:duct}.
A reference fully explicit numerical simulation has been carried out and used as a basis of reference
for the ATI algorithm, here applied to all coordinate directions. As seen in Table~\ref{tab:duct},
the corresponding CFL number is about unity.
As in the case of plane channel, DNS were carried out at increasing values of CFL, until deviations
from the reference data were found, to determine the maximum allowed time step for accuracy. 
It appears that accurate results of the semi-implicit algorithm are recovered
up to $\mathrm{CFL} \approx 10$. Again, implicit treatment of the $x$ direction is not capable of fully suppressing
the corresponding time step limitation, owing to the emergence of accuracy issues.
Similar to channel flow, use of the ATI algorithm allows for about $85\%$ cost reduction.
Figure \ref{fig:duct} confirms that excellent matching of the flow statistics is found
among DU02-ATI, DU02-EXPL and the data of \citet{pinelli_10}, except for some differences in
the wall-normal Reynolds stress and the pressure r.m.s., which may be due to the greater importance of acoustic waves
in the presence of a fully confined flow geometry.
%

\section{Conclusions} \label{sec:conclusions}

A novel semi-implicit algorithm for time-accurate solution of the compressible Navier-Stokes equations 
has been developed, which is capable to operate efficiently all the way from low
subsonic to supersonic flow conditions. The main features of the algorithm are as follows:
i) use of the entropy transport equation instead of total energy conservation;
ii) Beam-Warming-like linearization of the partial convective flux associated with acoustic propagation;
iii) energy-consistent discretization of the convective derivatives in the explicit part of the time-advancement operator;
iv) semi-implicit treatment of viscous fluxes based on isolation of Laplacian terms;
v) approximate factorization for implicit treatment of multiple space directions;
vi) third-order accurate Runge-Kutta time integration, according to the algorithm proposed by \citet{nikitin_06}.
The main advantage of the algorithm is that, unlike the classical Beam-Warming scheme, it avoids the computationally expensive
inversion of $5\times5$ block-banded matrices, but rather of standard banded matrices (tridiagonal matrices in the
case of second-order accurate space discretization). Specifically, a single banded matrix inversion is needed for 
implicit treatment of the convective terms, whereas five matrix inversions are needed if viscous 
terms are also handled implicitly. 
The cost overhead with respect to standard explicit algorithms 
(see Table~\ref{tab:cost}) is quite modest, ranging from $20\%$ to $30\%$, for each space direction to be handled implicitly.  
Modification of existing compressible flow solvers to incorporate the present method is straightforward, as the explicit 
part of the algorithm is unchanged.

The method nominally allows unconditional stability for low-Mach-number flows.
However, flux linearization and approximate factorization reduce the stability margins, 
and CFL number of the order of 5-10 are achieved in practical computations, 
which is probably less than achievable with iterative methods. 
However, compared to compressible flow algorithms based 
on pre-conditioning, the present method avoids use of inner time iterations, 
whose computational cost is difficult to estimate a-priori.
The other possible shortcoming of the method is the use of the entropy equation, which is instrumental
to achieve (approximate) separation of hydrodynamic from acoustic effects.
While use of the entropy equation yields improved numerical stability, it also makes 
proper capturing of shock waves difficult, as the equations are not in conservation form.
We have found that this issue can be fixed by locally reverting to a total energy formulation
for the explicit time increment in the presence of shocks, as identified through a shock sensor~\citep{pirozzoli_11a}.
The resulting time increments are then converted to the entropy increments, prior to application of 
the implicit operator.

Although the algorithm herein developed has in principle much wider range of applications, 
the main focus of this paper was on DNS of compressible wall-bounded flows,
which is notoriously plagued by severe time step restrictions inherited from the wall-normal acoustic and viscous
stability conditions. We have found that the wall-normal acoustic time limitation can be effectively 
removed through semi-implicit treatment. The same conclusion also applies to the viscous time step restriction, although
the most efficient way to remove it is placing the first grid point sufficiently away from the wall $y^+ \approx 0.5-0.7$,
and using suitable staggering~\citep{modesti_16}, with no effect of accuracy. 
The wall-parallel stability restrictions can also be suppressed through
semi-implicit treatment. However, accuracy considerations lead to the practical rule (see Fig.~\ref{fig:dtconv}) 
that the time step cannot be much larger than the one stemming from the streamwise time limitation. 
Hence, we suggest that in low-subsonic flow both the wall-normal and the spanwise convective terms are handled
implicitly, whereas the streamwise terms can be evaluated explicitly. The resulting saving of computer time
can then be of the order of $85\%$ with respect to a fully explicit solver.
In high subsonic or supersonic flow, implicit treatment of the wall-normal convective derivatives is 
sufficient, with typical savings of to order of $50\%$, in line with theoretical estimates.

We foresee that the present technique can be fruitfully extended to numerical simulation
of wall-bounded turbulent flows with time-accurate models, such as LES or DES~\citep{spalart_00}.
In that case, given the higher aspect ratio of near-wall cells, higher gains are expected.
Advantages with respect to classical algorithms based on Beam-Warming linearization are also
expected for steady RANS applications. Indeed, although the present algorithm is in principle 
only capable of suppressing the acoustic time step limitation, it is found to be at least as 
stable as Beam-Warming in practical computations.\\
\\
{\bf Acknowledgements}\\
We acknowledge that most of the results reported in this paper have been achieved using the PRACE Research Infrastructure resource FERMI based at CINECA, Casalecchio di Reno, Italy.

\bibliographystyle{elsart-num-names}
\bibliography{references} 

\end{document}